\documentclass[number,preprint,12pt]{elsarticle}

\usepackage{amssymb}

\usepackage[numbers]{natbib}
\usepackage{hyperref}
\hypersetup{colorlinks,allcolors=black}
\usepackage{subfigure}
\usepackage{threeparttable}
\usepackage{booktabs}
\usepackage{amsmath}
\usepackage{float}
\usepackage{fancyhdr}

\newcommand*{\doi}[1]{DOI \href{https://doi.org/#1}{\texttt{#1}}}
\makeatletter
\def\bibinfo#1{%
  \@ifundefined{bibinfo@X@#1}%
    {\@firstofone}
    {\csname bibinfo@X@#1\endcsname}}
\makeatother

\fancypagestyle{pprintTitle}{%
\lhead{\small \textbf{This article has been submitted to Elsevier for possible publication. Copyright may be transferred without notice and this version might not be longer available}\\} \chead{}\rhead{\scriptsize}
\lfoot{\small \textit{Preprint submitted to Computer Networks}}\cfoot{}\rfoot{\small \textit{\today}}

\setlength{\headheight}{52.4842pt}
\addtolength{\topmargin}{-20.4842pt}
}

\journal{Computer Networks}

\begin{document}

\begin{frontmatter}



\title{ML-powered KQI estimation for XR services. A case study on 360-Video}

\affiliation[inst1]{organization={Telecommunication Research Institute (TELMA), Universidad de Málaga},
            addressline={E.T.S. Ingeniería de Telecomunicación, Bulevar Louis Pasteur 35}, 
            city={Malaga},
            postcode={29010}, 
            state={Andalucia},
            country={Spain}}

\author[inst1]{O. S. Peñaherrera-Pulla}
\ead{sppulla@ic.uma.es}
\author[inst1]{Carlos Baena}
\ead{jcbga@ic.uma.es}
\author[inst1]{Sergio Fortes\corref{cor1}}
\ead{sfr@ic.uma.es}
\author[inst1]{Raquel Barco}
\ead{rbm@ic.uma.es}

\cortext[cor1]{ Corresponding author}
\fntext[label2]{ This work has been partially funded by: Ministerio de Asuntos Económicos y Transformación Digital and European Union - NextGenerationEU within the framework ``Recuperación, Transformación y Resiliencia y el Mecanismo de Recuperación y Resiliencia'' under the project MAORI, and Universidad de Málaga through the ``II Plan Propio de Investigación, Transferencia y Divulgación Científica''. 
This work has been also supported by Junta de Andalucía through Secretaría General de Universidades, Investigación y Tecnología with the predoctoral grant (Ref. PREDOC\_01712) as well as by Ministerio de Ciencia y Tecnología through grant FPU19/04468.}

\begin{abstract}

The arise of cutting-edge technologies and services such as XR promise to change the concepts of how day-to-day things are done. At the same time, the appearance of modern and decentralized architectures approaches has given birth to a new generation of mobile networks such as 5G, as well as outlining the roadmap for B5G and posterior. These networks are expected to be the enablers for bringing to life the Metaverse and other futuristic approaches. In this sense, this work presents an ML-based (Machine Learning) framework that allows the estimation of service Key Quality Indicators (KQIs). For this, only information reachable to operators is required, such as statistics and configuration parameters from these networks. This strategy prevents operators from avoiding intrusion into the user data and guaranteeing privacy. To test this proposal, 360-Video has been selected as a use case of Virtual Reality (VR), from which specific KQIs are estimated such as video resolution, frame rate, initial startup time, throughput, and latency, among others. To select the best model for each KQI, a search grid with a cross-validation strategy has been used to determine the best hyperparameter tuning. To boost the creation of each KQI model, feature engineering techniques together with cross-validation strategies have been used. The performance is assessed using MAE (Mean Average Error) and the prediction time. The outcomes point out that KNR (K-Near Neighbors) and RF (Random Forest) are the best algorithms in combination with Feature Selection techniques. Likewise, this work will help as a baseline for E2E-Quality-of-Experience-based network management working in conjunction with network slicing, virtualization, and MEC, among other enabler technologies.

\end{abstract}


\end{frontmatter}
\thispagestyle{fancy} 

\section{Introduction}
\label{sec:Introduction}

The new generation of services aims to revolutionize our day-to-day activities as well as the way people interact with others. These novel services that involve cutting-edge technologies like Extended Reality (XR) are intended to bring different levels of virtual and enriched experiences to our lives. Although the "virtual" topic has been discussed since decades ago, the enabler technologies were not ready as they are nowadays.

In this context, the implications of XR in the real life are going to be omnipresent in all tasks and human activities. For instance, the metaverse concept, recently reinvigorated by the META company \cite{Meta2022CompanyMeta}, looks to bring physical human interactions (e.g. meetings, entertainment, shopping, ...) to the virtual world in a real-like manner.  This fact, in concordance with the development of new radio mobile technologies, has aroused a hot topic that must be approached in order to integrate this kind of service into the network. This perspective will generate new exploitation ways of these features by vertical vendors and network operators in an organized and standardized manner. 

In general terms, the XR is cataloged into different sub-technologies that represent the level of virtualization of reality. On one side, Augmented Reality (AR) is focused on overlying virtual elements (e.g. information, rendered objects, etc.) to interact with physical reality. To reach this, the physical reality is captured and processed in order to generate models that feedback on the experience. On another side, Virtual Reality (VR) aims to generate a whole alternative experience, where every element is generated virtually, rendered, and displayed to the user. However, information from the physical information is required (i.e., user tracking). The mixture of both technologies is known as Mixed Reality (MR) whose main key is to introduce physical and virtual features at the same time, this way deploying a different degree of immersion (e.g. a real human avatar inside a 3D and fully virtualized environment) \cite{Wang2022APrivacy}.

In light of VR technology, several applications of XR are drawing attention in the research world. One of them is 360-degree Video (or simply 360-Video) that intends to create immersive experiences for the user. The basic concept of this application is to deliver multimedia content in an omnidirectional manner controlled by intuitive human-based actions, this way the user can feel inside the video. To reach this goal a Head-Mounted Device (HMD) should be used. Given that, various "traditional" video providers like YouTube are presenting alternatives to enjoy 360-Video on their platforms \cite{YouTube2022YouTubeVR} using HMDs, computers, tablets, etc.

Despite the fact that some service providers are working on this topic, delivering the content from the servers to the user equipment is not trivial. In this sense, VR technology requires some specific conditions to properly perform. The lack of resources can lead to not just a bad user experience, but also to some physical problems such as cybersickness \cite{Gavgani2018ADifferent}, confusion, anxiety, fatigue and even physical injuries \cite{LaMotte2017VirtualCNN, Chang2020VirtualMeasurements, Costello1997HealthLiterature}.

Previous studies have demonstrated that to reach a real-feel 360-Video experience, the content should be provided with a minimum resolution of 60 pixels per degree at a recommended 120-Hz frame rate \cite{Elbamby2018TowardReality}. Moreover, the influence of the startup time of the video and the quantity of stalling events can decimate the quality of the experience. To overcome all these barriers, mobile technologies like 5G and Beyond-5G (B5G) are being designed and developed using different architectural concepts in comparison with LTE or other legacy networks. 

The new-generation mobile networks aspire to convert new services into native services that make use of the network and computational resources based on their requirements. This approach will provide the networks with mechanisms that ensure proper levels of quality for each service based on automatic and intelligent resource policies. For this purpose, different enabler technologies and features of 5G and 6G will be used such as Network Slicing, Network Functions Virtualization (NFV), SDN (Software-Defined Networks), SDR (Software-Defined Radio), MEC (Mobile Edge Computing), and AI/ML (Artificial Intelligence / Machine Learning) \cite{Mahmoud20216G:Problems}. Another important concept that is drawing attention from researchers, operators, and vertical vendors is the Open RAN trend \cite{O-RANAlliancee.V.2022O-RANUs}, which is intended to deploy fully intelligent, virtualized, and interoperable mobile networks 

The aforementioned features are in the development stage and they are not directly applicable to the legacy networks, where all the resources are treated based on different policies that are manually tuned by the operators. This fact lies in that most of the services remain as data that goes across the network in an OTT (Over-The-Top) fashion. Nonetheless, new-generation multimedia services will be intended to reinvent the manner in which operators provide connectivity, users digest content and vertical vendors exploit cutting-edge technologies \cite{Laurell2019ExploringTrialability}. All these parties will be involved in different application fields ranging from education, healthcare to military or entertainment \cite{Rahimizhian2020, Ahir2019ApplicationSports}.

The key idea that motivates this work is the use of performance metrics and network configuration parameters (e.g. Key Performance Indicators - KPIs) to estimate Key Quality Indicators (KQIs). This supposes a suitable strategy that enables the operators to manage the network based on objective metrics instead of subjective ones using MOS (Mean Opinion Score) or user-assisted metrics. However, the main disadvantage of subjective metrics is the high correlation that they have with the service since they are based on their perceptions that may be affected by the system (content/media related), context (physical, temporal even social) or human (physical, emotional, ...) impact factors (IFs) \cite{Zhao2017QoEStrategy}. Despite this, the use of objective metrics extracted from network information, reachable by operators, may help to reduce CAPEX and OPEX costs. Likewise, it may enhance the network robustness in order to provide high-quality services tuned by a user-centric perspective.

To the best of the authors' knowledge, there is no previous research that involves the study of ML algorithms to estimate the metrics of quality for the 360-Video case. Most previous work has been developed on the estimation of subjective metrics \cite{Anwar2020, Yao2019TowardsReality}, image quality \cite{Hanhart2018, Tran2017} or researched on improvements related to the streaming methodology \cite{Park2020} or playout estimation using emulated scenarios \cite{fihlo2019} for 360-Video service.

Hence, the key contribution of this work focuses on the development of a novel ML-powered framework to estimate the Quality of Experience (QoE) based on the use of KQIs. This framework exploits the information self-contained in the network information such as metrics, statistics, and configuration parameters (e.g KPIs) to enhance the management of service-oriented new-generation mobile networks. To reach this, several ML algorithms have been integrated, trained, and tested in order to foresee the values of high-level 360-Video metrics such as resolution, frame rate, initial playback time, average stall time, and client-side throughput among various others. For training purposes, a dataset has been generated using a testbed \cite{Penaherrera-Pulla2022KQI5G}. This allows for assessing the performance of different automated experiments, as well as establishing various scenarios with variable transmission power, noise, and channel bandwidth conditions. Obtained results are used to train multiple ML algorithms using a search grid approach in order to find the best models through hyperparameter tuning.    

In this way, this paper is organized as follows. First, Section \ref{section:Related} provides a brief outlook of the state of the art related to 360-video and ML approaches for QoE. Then, Section \ref{section:Framework} describes the framework adopted to generate a dataset for training and testing procedures. Additionally, the section provides a comprehensive description of the design and implementation of ML mechanisms, which are the cornerstone for the service's KQI estimation. After that, in Section \ref{section:Evaluation} the algorithms are assessed to establish which one performs the best for each parameter in two scenarios, a per-sample estimation and for a session-average estimation. Finally, Section 5 provides the conclusions of the work, exposing the key points of this research as well as outlining some future work.

\section{Related work}

\label{section:Related}

\subsection{KQI estimation}

KQI estimation is considered one of the most important current strategies to objectively manage mobile networks. Several services have been used as study cases such as traditional video streaming, FTP, HTTP, and so on. This approach is suitable to manage correctly 5G and B5G networks in order to guarantee proper quality service levels. Moreover, it is useful for supporting the correct resource management in an automated perspective using only network information that is well-known and reachable to the operators.

In the same context of KQI estimation, Herrera et al. \cite{Herrera-Garcia2019Modeling5G} describe their work as a methodology to meet the service performance through the use of KPIs which depicts the network performance and behavior. With this approach, the network operator is able to estimate an objective perspective of the user's experience without the need to trespass the level of intrusiveness as other methodologies do, for instance, packet inspection. This work offered a mechanism to estimate KQIs for FTP service.  

Baena et al. proposed in \cite{Baena2020EstimationScenarios} an approach to estimate KQIs in a Network Slicing scenario for a video streaming service. The metrics are estimated from network information and statistics. This approach is useful in the context of the new-generation networks, where the operators need to know the quality perceived by the user but also use this information to estimate possible resources required and their pricing. 

Conversely, Wassermann et al. \cite{Wassermann2020ViCryptTraffic} describe a different approach for KQI estimations for HAS (HTTP Adaptive-video Streaming). This strategy infers stalling, resolution, and throughput based on mechanisms that use estimation and classification techniques. The key contribution of this work is the use of pure network metrics such as packet-level statistics. However, its application is limited to the protocols and data patterns for HAS.

In this setting, Zhang et al. \cite{Zhang2021UserNetwork} proffer a KPI-driven KQI mapping based on qualitative levels. The authors present an Adaptive Naive Bayesian Classifier, and compare it with KNN (K-Near Neighbors) and Gaussian Kernel Function, to establish the state (ranging from unacceptable to excellent) of KQIs for video, IM (Instant Messaging), and web services. The results are assessed through accuracy and various specificity metrics.

Despite these proposals, there is no prior research done in the estimation of objective metrics for 360-Video or any related applications belonging to XR technologies, which highlights the contribution of our work.

\subsection{ML for QoE}

Offering a high-quality video service is one of the main objectives for operators and service providers at the current times. In this scope, ML has been introduced as a useful tool for improving the Quality of Services (QoS). Different applications are mentioned in the bibliography such as the prediction of the head movements of the user, the spatial encoding of video using tiling methodologies, as well as the identification of traffic, different from packet inspection strategies, or MOS modeling, among various others.

Some previous studies have been done to establish, numerically, the relations between service engagement and quality. In \cite{Dobrian2013UnderstandingEngagement} the authors show the strong correlation of high-level view engagements with low startup times, low buffering times, low rebuffering number of events and a considerable high resolution.

The authors in \cite{BenLetaifa2018RealServices} present an ML-based mechanism to estimate the QoE through MOS. The outcome models were intended to calculate the subjective QoE using metrics such as PSNR (Peak Signal to Noise Ratio), bitrate, throughput, and VQM (Video Quality Metric), among various others. The algorithms were trained using a dataset that gathered the people´s assessment of the video quality using a testbed.  

A different application of ML for QoE is analyzed in \cite{vanderHooft2015AClients} where the authors present a strategy aiming to increase the QoE. The quality of the experience is assessed through the MOS of the video service based on an ML mechanism that manages the adaptive streaming. This approach uses the bitrate of the link to handle the buffer filling time, in this way improving the QoE. Moreover, in  \cite{Jahromi2018TowardsSDN} an ML approach is developed to characterize the QoE of an HTML service through KPIs using a testbed that exploits SDN flexibility. The KPIs (e.g. bandwidth, TX (transmission), and RX (reception) load, delay, etc.) are estimated based on the network information gathered in several experiments.

Rothenberg et al. \cite{Rothenberg2020Intent-basedEstimation} propose an ML approach to manage decision-making in the context of DASH video streaming using SDN. This work is based on the use of ML to map the MOS from the KPIs of the network. Then, an orchestrator decides which high-level policy should be taken into account by network elements to manage the policies and strategies (e.g. routing). The data is gathered using a collector that develops traffic mirroring for processing information in a MEC. Nevertheless, mirroring traffic (traffic duplication) is becoming ineffective for network operators. A different approach is presented by Gutterman et al. \cite{Gutterman2019Requet:Traffic}, where the QoE estimation for the service, particularly YouTube video service, is done through an ML-based algorithm whose inputs are statistics extracted from IP headers.

Furthermore, a related topic is proffered by Kattadige et al. \cite{Kattadige2021360NorVic:Traffic}, where an ML mechanism to identify 360-Video traffic from YouTube and Facebook using encrypted packet-level data is described. This approach in conjunction with additional information may be applied to understand the network behavior under specific kind of traffic. In this setting, the recognition of certain patterns may enhance certain dedicated tasks such as codification or caching strategies at the operator or content provider sides.

In the same context, Begluk et al. \cite{Begluk2018MachineNetwork} present a MOS estimation scheme for video streaming services. In their proposal, the MOS is estimated through an ANN (Artificial Neural Network) whose inputs are typical QoS metrics such as delay, jitter, and packet loss. The original MOS values were gathered by testing people using a mobile phone using an emulated LTE network.

Concluding, there is no previous related work oriented to either the objective estimation of the QoE on XR or the specific 360-Video use case proposals. In this context, the key contribution of this work is the introduction of a novel ML-powered framework that aims to ease the dynamic management of the network based on information available for the own network from a service QoE perspective.

\section{Framework}

\label{section:Framework}

In this section the methodology applied is described. This consist of a series of diverse ML algorithms, which have been trained and tuned, using a k-folding and cross-validation approach, to estimate KQIs for the 360-Video system. To reach this objective, the input considers a dataset, made of KQIs, and metrics and statistics from the network, that has been previously obtained using a testbed to generate multiple iterative tests under different channel conditions.

Regarding the architecture of the service, the client side is performed by an HMD (Head Mounted Device) and a CPE (Customer Premises Equipment). The first one is intended to display the 360-Video content to the user and to collect KQIs, through a dedicated application developed in Unity 3D. This application allows displaying the content while metrics are being gathered in the background. In addition, the processing and rendering tasks are executed integrally using the HMD's hardware due to the implementation of a standalone architecture. Differently, the CPE is used as a bridge between the mobile network and the WiFi HMD access. Furthermore, some network performance metrics are collected in this device as well as in the transport network. 

The transport network is featured by a Crowdcell device, which is an open-source solution that mixes SDR with a virtualized core, in this way, acting like a whole network-in-a-box \cite{Baena2021ASolutions}. In the same context, this framework allows the emulation of some radio impairments such as attenuation and noise presence due to the use of the SDR module. Additionally from the functionalities mentioned, this solution provides some metrics related to radio performance that are included in the input dataset. 

\begin{figure}[ht]
    \centering
    \includegraphics[trim = 0 0 0 0, clip, scale = 0.5]{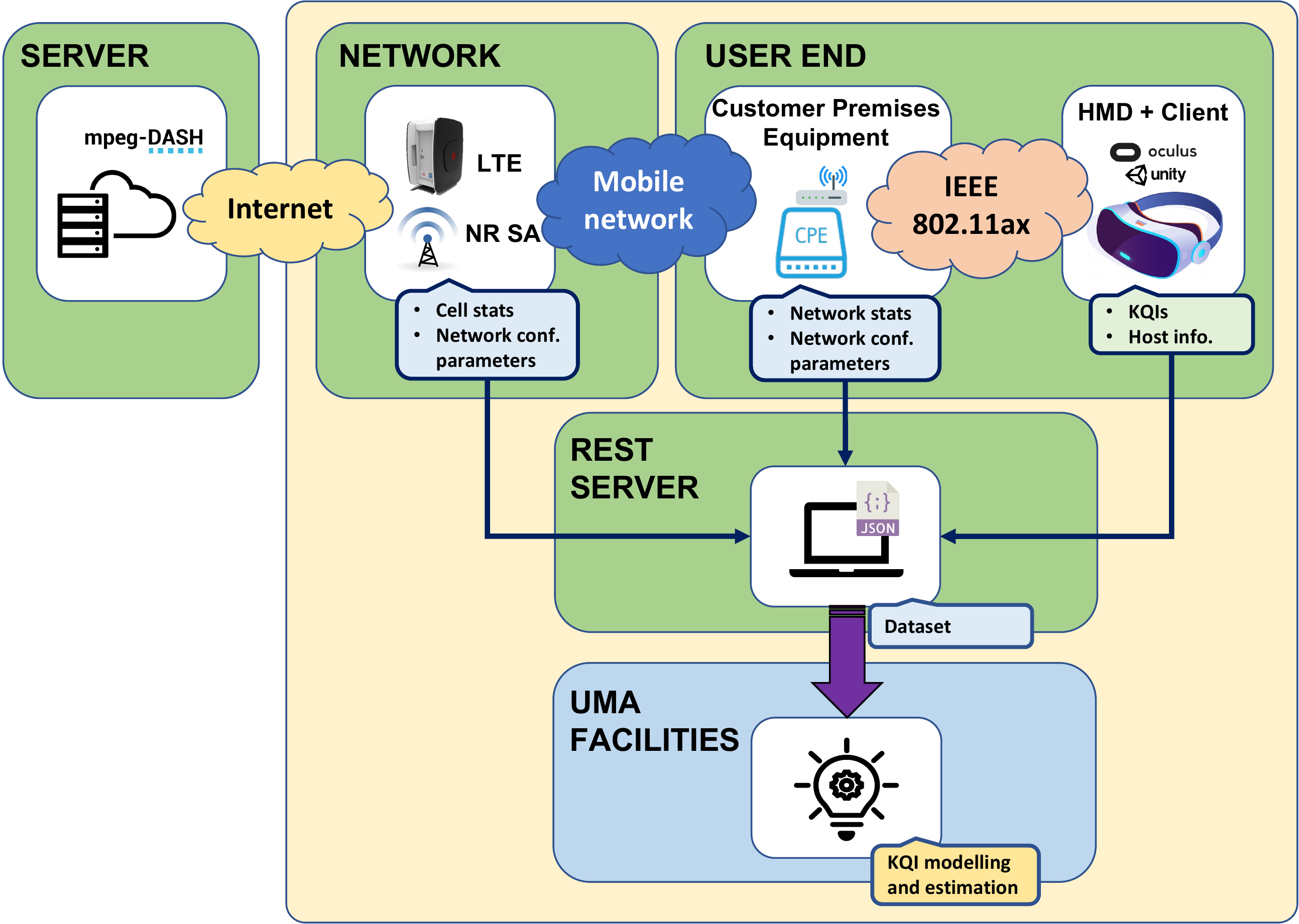}
    \caption{Implementation scheme.}
    \label{fig:SetupTestbed}
\end{figure}

In addition to these elements, a RESTful (Representational State Transfer) server was implemented to serve as the storing point of the measurements done on the client side as well as in the network. This scheme is depicted in Figure \ref{fig:SetupTestbed}.

\subsection{Dataset acquisition}
\label{section:dataset}

In order to acquire the dataset that serves as the input for the ML models, several experiments \cite{Penaherrera-Pulla2022KQI5G} were done using the testbed depicted in Figure \ref{fig:SetupTestbed}. The experiments were intended to display 360-Video in an iterative manner, this way assuring that all the tests use the same multimedia content and guarantee objectiveness. Then, the results gathered captured the network influence over the service through several configurations as follows in Tables \ref{table:TestbedConfCrowd} and \ref{table:TestbedConfMedia}.

\begin{table*}[ht]
\begin{threeparttable}
    \caption{Testbed configuration - Crowdcell experiments setup.}
    \label{table:TestbedConfCrowd}
    \centering
    \begin{tabular}{p{0.2\textwidth}p{0.5\textwidth}p{0.2\textwidth}}
    \hline
    \textbf{Parameter} & \textbf{Description}& \textbf{Value} \\\hline
    Experiment duration & Duration of video playback in seconds & 120\\
    Sampling frequency & Number of samples per second & 1\\
    Experiments & Number of experiments with the same configuration & 60\\
    Technologies & Transport network technologies connecting the user equipment with the Internet & 4G/LTE \par 5G\\
    Crowd-BW & LTE Crowdcell channel Bandwidth & $5$ MHz \par $10$ MHz \par $15$ MHz \par $20$ MHz\\
    \hline
    MaxPT~$^{1}$ & Crowdcell maximum power transmission level for \textit{MaxPT} tests & 0 dB\\
    MinPT~$^{1}$ & Crowdcell minimum power transmission level for \textit{MinPT} tests & -10 dB\\
    RedPT~$^{1}$ & Crowdcell power transmission level for \textit{RedPT + Noise} tests & -20 dB\\
    Max-Noise~$^{1}$ & Maximum noise level for \textit{RedPT + Noise} tests & -20 dB\\
    Min-Noise~$^{1}$ & Minimum noise level for \textit{MaxPT and MinPT} tests & -30 dB\\
    \hline
    
    \end{tabular}
    \begin{tablenotes}
      \item[1]{Device-internal configuration parameter.}
    \end{tablenotes}

\end{threeparttable}
\end{table*}

\begin{table*}[ht!]
\begin{threeparttable}
    \caption{Testbed configuration - Media.}
    \label{table:TestbedConfMedia}
    \centering
    \begin{tabular}{p{0.2\textwidth}p{0.5\textwidth}p{0.2\textwidth}}
    \hline
    \textbf{Parameter} & \textbf{Description}& \textbf{Value} \\\hline
    
    Video resolution & Video available resolutions at the server & $720\times360$ \par $1080\times540$ \par $1440\times720$ \par $2160\times1080$ \par $3840\times1920$\\
    Average bitrate per segment & Average bitrate per each video segment (same order as resolutions) & 1 Mbps \par 1.5 Mbps \par 3 Mbps \par 5 Mbps \par 9 Mbps\\
    Frame rate & Frame rate at which video is encoded & 30 FPS\\
    Segment duration & Time period for each video segment & 4 seconds\\
    Codec & Video codec used & avc1.42c00d\\
    Initial buffer & Filling time for initial playback & 5000 ms\\
    Min. \& Max. buffer threshold & Minimum and maximum thresholds for buffer & 50000 ms\\
    Streaming protocol & Protocol used for media streaming & Standard DASH\\
    ABR strategy & Adaptive Bitrate strategy used for buffer filling & Throughput-based\\
    \hline
    HMD Model & HMD model used in the testbed & Oculus Quest 2\\
    Operating System & HMD operating system & Android OS 10 / API-29\\
    Graphic engine & Graphic engine that supports the video 360 client & Unity 2020.2\\
    Video player & Video player API integrated into the video client & ExoPlayer NonOES\\
    \end{tabular}
\end{threeparttable}
\end{table*}

\begin{figure}[t!]
    \centering
    \includegraphics[trim = 0 5 10 0, clip, scale = 0.5]{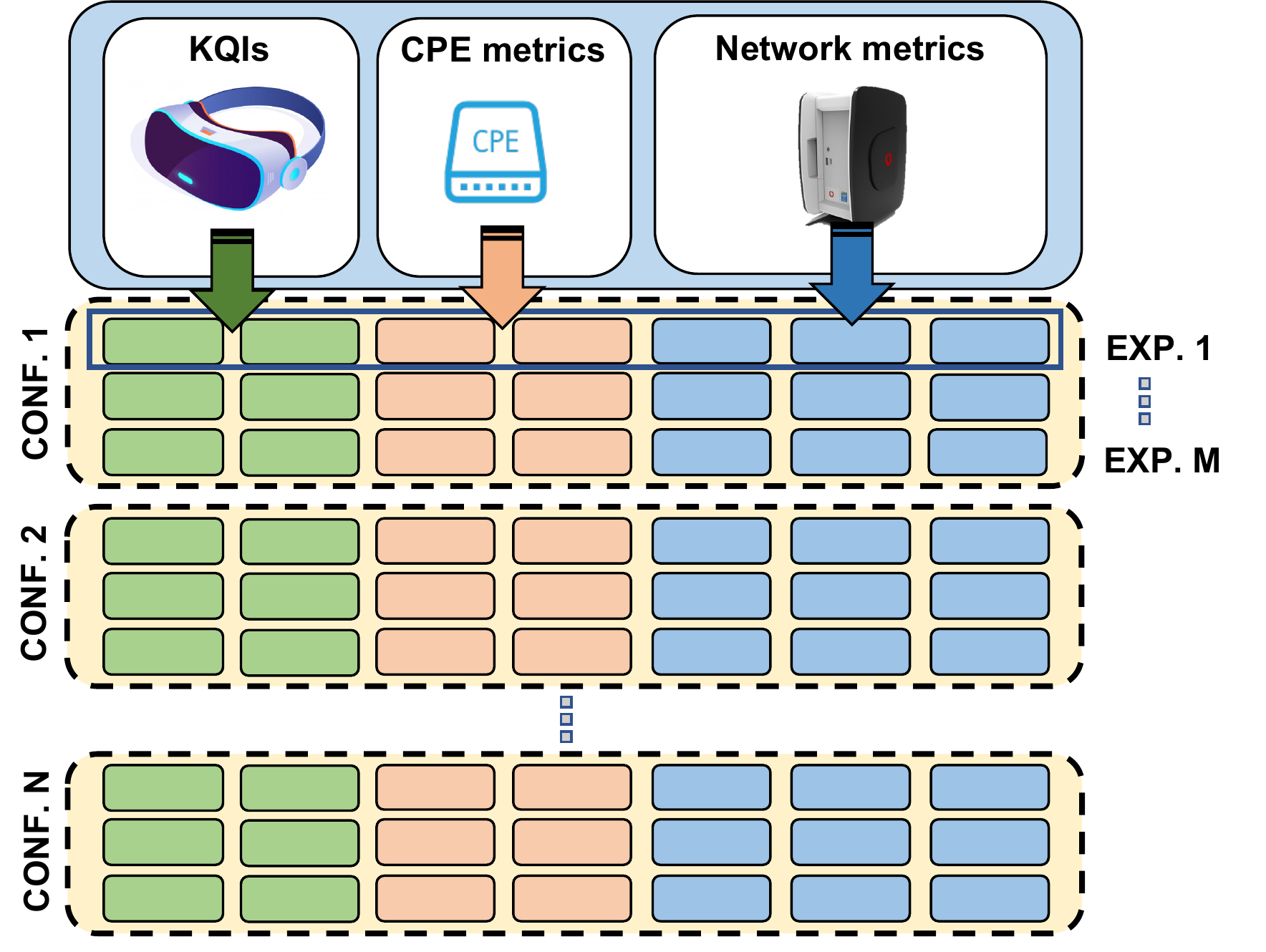}
    \caption{Dataset creation.}
    \label{fig:Dataset}
\end{figure}

The work methodology consists of 12 different configurations of the transport network. Each one is composed of 120-minute-long experiments where samples are obtained for each second of video displayed. Conversely, the CrowdCell provides LTE access as well as generating different channel conditions generated by transmission power changes, channel bandwidth and noise emulation using the SDR module. Then, the REST server gathers the metrics obtained in the HMD as well as the ones fetched by the Crowdcell and CPE, this way generating an integral dataset that represents the service performance from a high-level perspective as well as from a network viewpoint. The interpretation of this process can be seen in Figure \ref{fig:Dataset}.

With regard to the service metrics, some recommended KQIs for 360-Video are measured in the user equipment such as resolution, frame rate, initial playing time, stall events, throughput, and latency, among others. Likewise, on the network side, some metrics collected by the Crowdcell and CPE are basically counters and KPIs of the network and configuration parameters such as channel bandwidth, carrier frequency, throughput, number of retransmissions in uplink as well in the downlink direction, and so on and so forth.

\subsection{Model training}
\label{section:training}

The dataset generated in the previous subsection is used to train ML algorithms and produce the models that describe the service performance from an E2E (End-to-end) perspective using KQIs. These models are based on the information contained in KPIs and network stats and configuration parameters (e.g. covariance, non-linear dependency) that reflect their influence on the experience. The implemented mechanisms are able to estimate several performance metrics such as resolution, frame rate, initial playback time, stall events, latency, and throughput, among others.

The proposed framework consists of a software pipeline designed to foster several stages or procedures in an organized manner. This methodology is assumed in order to assure all the processes are done in the right order (transformations, training, and posterior assessment) but also avoid possible statistical leaking of the test data into the train data, this way guaranteeing the objectiveness of the training phase respect to its testing one.

Before the pipeline is applied, the dataset went through a two-step preprocessing phase to delete all the samples that contain measurement errors (invalid values due to external influence) or certain experiments that suffered issues during their evaluation (disconnection with the radio cell, electrical, or processing outages and so on) in its first step. The later stage consisted of deleting all the parameters or variables (since here called estimators) where their variance is zero, which means that have no variation along every experiment executed. These variables generally featured textual information or network or client configurations that were maintained constant during the testing.

It is important to mention that the original dataset was composed of a total of 86400 samples. This is the result of the multiplication of the number of radio channel bandwidths (4) by the power transmission scenarios (3), the number of samples per experiment (120-second experiment with a sampling frequency of 1~Sample/s), and the number of experiments for each configuration (60). After the two-step dataset preprocessing, the dataset is organized in an experiment-average format which means calculating the average predictors per session (720 experiments).

Previously to the training phase, the dataset has been subject to standardization, so that its scales are modified to be adequately used in the training of the algorithms. Likewise, the mean of each metric has been eliminated, thus allowing obtaining indicators that vary positively or negatively around 0, this value being possible when the metric is at its mean.

For experimental purposes, the framework has been designed in order to test three scenarios: (I) estimation with no feature engineering techniques, (ii) feature selection (FS) using an importance ranking, and (iii) prediction of KQIs using feature extraction (FE) using PCA (Principal Component Analysis). It is remarkable to point out that the dataset has been previously split into a training and a testing set. The former split is used to train and tune the estimation model while the latter allows for validating the model performance. The splitting strategy adopted has been 70\% and 30\%.

The first case is the lowest-complexity strategy of estimation. This consists of inputting the standardized dataset into the pipeline, with no extraction or selection stages (i.e., neither PCA nor another feature engineering strategy). This is done in order to check if no previous data treatment is needed according to the nature of the collected data. Despite this, the next stage, the model training process, in the pipeline is common for the three scenarios.

The second case involves a feature selection methodology featured by a \textit{SelectFromModel} strategy from Sciky-learn \cite{Pedregosa2011Scikit-learn:Python}. This methodology allows the algorithm to input only the best predictors that remarkably impact the KQI estimation. however, it may incur an increase in the processing times due to the intermediate stages in the training pipeline. These best predictors are obtained through the application of a Random-Forest-based technique, which offers a sort list with the highest impact predictors.

With respect to the third case, a PCA stage is used for feature extraction. This consists of mathematically transforming and separating the original information into certain key-information components (known as Principal Components). Here, the closer is the secondary components from the first one, the more variability of the dataset is captured. Applying this methodology, it is possible to generate an output dataset that synthesizes the original information (patterns, statistics, correlation between variables) into new and simpler predictors. In the end, the best components involving $\>=95\%$ are extracted. In addition, this procedure is applied separately for the train and test sets.

\begin{figure}[t!]
    \centering
    \includegraphics[trim = 0 0 0 0, clip, scale = 0.5]{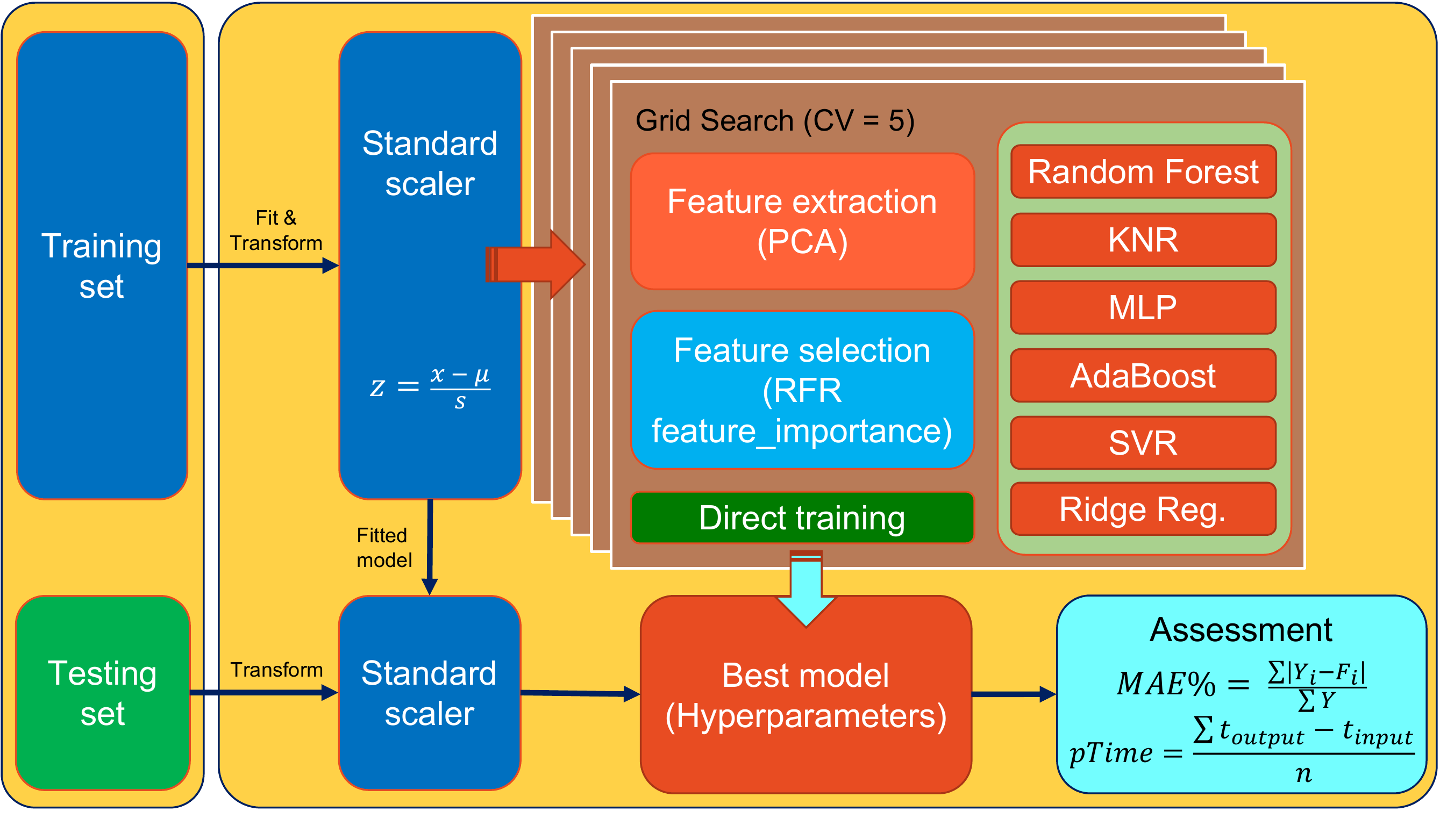}
    \caption{ML approach}
    \label{fig:ML_approach}
\end{figure}

To establish the best model for each algorithm, a search strategy was used to find the best hyperparameters that achieve the best performance with the validation set. For this purpose, the \textit{Grid Search} algorithm with a 5-fold CV (cross-validation) strategy \cite{Pedregosa2011Scikit-learn:Python} has been used. This approach is intended to split the whole training set into several subsets (in this case 5) to train the model with certain conditions determined by a group of predefined hyperparameters that are passed to the algorithm. This training process is repeated with each subset, while the cross-validation subset determines which hyperparameter configuration performs better. In this way, the use of this technique usually leads to model overfitting avoidance, boosting the performance of the models in different scenarios. An overview of this approach is depicted in Figure \ref{fig:ML_approach}.

The algorithms considered in the framework are \cite{Pedregosa2011Scikit-learn:Python}:

\begin{itemize}
    \item Random Forest (RF).
    \item Ridge Regression (RR).
    \item Support Vector Machine Regression (SVR).
    \item K-Neighbors Regressor (KNR).
    \item Multilayer Neural Networks (Perceptron - NN). 
    \item AdaBoostRegressor (ABR).
\end{itemize}

To evaluate the performance of the algorithms the test set has been inputted to the model that performed the best with the specific hyperparameter configuration determined by the training phase. This allows calculating some metrics based on how the model is able to estimate the KQIs. The metrics that have been taken into account for the quantification of the feasibility of each algorithm are MAE (Mean Absolute Error), and the prediction time.

In this sense, it is well known that R2 metrics are not the best option to assess models due to their linear dependency. MSE (Mean Squared Error) and RMSE (Root Mean Squared Error) do not provide enough information about the overall trend but its variation around the mean. Likewise, MAPE (Mean Average Percentage Error) may lead to erroneous interpretations of the performance when values are close to zero. In addition, a scaled version of MAE is used in the discussion of the results in Section \ref{section:Evaluation}. This modification allows converting the absolute scope of the MAE into a relative scale (0 - 1), which eases the analysis and comparison between KQIs of different units or scales, for instance, resolution in pixels or latency in milliseconds. The selected assessment metrics can be described as follows:

\begin{itemize}
    \item{MAE - (Mean Absolute Error): This metric refers to the average error of the model obtained through the sum of each individual absolute error accounted for within the testing set.
    
    \begin{equation}
        \centering
        MAE = \frac{1}{n} \sum e_{forecast}
    \end{equation}}  
    
    \item MAE\% - (Scaled Mean Absolute Error): This indicator is a modification of the traditional MAE into a percent version through escalation using the mean of the real value of the ground truth. In this context, the closer the value is to zero, the best estimation is obtained.
    
    \begin{equation}
        \centering
        MAE\% = \frac{\frac{1}{n} \sum e_{forecast}}{\frac{1}{n}\sum y_{real}} = \frac{\sum e_{forecast}}{\sum y_{real}}
    \end{equation}
    
    \item $pTime$ - (Average prediction time): This metric defines the average time required for the pre-trained model to estimate the output for a specific KQI. The prediction time is accounted as the period between the predictors are inputted and the estimation output is generated.
    
    \begin{equation}
        \centering
        pTime\% = \frac{\sum t_{output} - t_{input}}{n}
    \end{equation}
    
    
    

\end{itemize}

Furthermore, the grid of parameters that were evaluated using this ML approach depends on the type of algorithm used. In order to assure algorithm convergence and affordable training times, some values have been selected by trial and error tests. These value ranges have been previously tested individually for each edge in order to assure that its value is valid for each algorithm. The aforementioned values are summarized in Table \ref{table:MLOptimization}.

\begin{table*}[ht!]
    \caption{ML hyperparameter optimization.}
    \label{table:MLOptimization}
    \begin{tabular}{p{0.25\textwidth}p{0.28\textwidth}p{0.40\textwidth}}
    \hline
    \textbf{Algorithm} & \textbf{Hyperparameter} & \textbf{Values} \\\hline
    Random Forest & n\_estimators \par max\_depth & [1, 2, 3, 4, 5, 6] \par[5, 6, 7, 8, 9, 10]\\
    \hline
    Ridge Regressor & alpha\par fit\_intercept & [$10e^{-5}, ..., 10e^5$] \par[false, true]\\
    \hline
    SVM Regressor & kernel \par degree \par $\epsilon$ \par C & [poly, rbf, sigmoid] \par [1, 2, 3, 4, 5, 6, 7] \par [0.01, 0.1, 0.5, 1.0] \par [0.1, 1, 10, 100]\\
    \hline
    K-Neighbors Regressor & leaf\_size \par n\_neighbors \par p & [10, 20, 30] \par [2, 4, 6] \par [1, 2, 3]\\
    \hline
    Perceptron & alpha \par hidden\_layer\_sizes & [0.0001, 0.0003, 0.001, 0.003, 0.01] \par [(80,), (100,), (80, 80), (100, 100), (80, 80, 80), (100, 100, 100), (200, 200, 200)]\\
    \hline
    AdaBoost Regressor & n\_estimators \par learning\_rate & [50, 75, 100, 125, 150] \par [0.0, 0.333, 0.666, 1.0]\\
    
    \hline
    
    \end{tabular}
\end{table*}

\section{Evaluation}
\label{section:Evaluation}

In this section, the results obtained throughout a campaign of diverse tests are described and discussed. The outcomes are intended to represent the performance of the different KQI estimation models for the average KQIs per session. For this purpose, the samples in the original dataset were grouped in experiments of 120 samples (i.e., representing a session of 120 seconds) and averaged according to their nature (e.g. median value of resolution because this parameter uses only integer values, frame rate uses mean values, etc.). 

Even though the use of fewer samples in the training set may affect the prediction accuracy, it can bring some advantages. One of them is the reduction in the estimation times incurred by the algorithms as well as the model complexity. Supporting this idea is the fact that actual networks are not able to continuously modify their configuration parameters within very short periods (in order of seconds) for a specific service. So, considering an estimator with a resolution time of a couple of seconds (i.e., the duration of a short session of video) can be a useful baseline for future optimization implementation.  

In this context, the algorithms, pointed out in the previous section, have been trained (using as predictors only CPE and Crowdcell network statistics and configuration parameters), tuned, and evaluated using the pipeline according to the framework described in Section \ref{section:Framework}. The results discussed in this section correspond to the MAE\% and the estimation time for different KQIs. The estimated metrics considered are video resolution, frame rate, initial startup/playing time, stalling time, throughput at the client side, and latency. 

Additionally, the nomenclature used will be standard throughout this section, this way, \textit{No\_FE} stands for Non-feature-engineering approaches \textit{FS} for Feature Selection and \textit{FE} for Feature Extraction.

According to the performance assessment basis, the closer the value of MAE\% is to zero, the best estimation is achieved. For that reason, the best-performing algorithm produces the lowest value of error. For the discussion of the results, MAE\% values lower than 10\% will be considered proper estimations. Values between 10\% and 20\% will be established as suitable estimations. Likewise, higher MAE\% values until 50\% are acceptable, meanwhile, the ones higher than that threshold will be labeled as inappropriate.

\begin{figure*}[ht]
\centering
\subfigure[Estimation error]{
    \includegraphics[width = 0.47\textwidth]{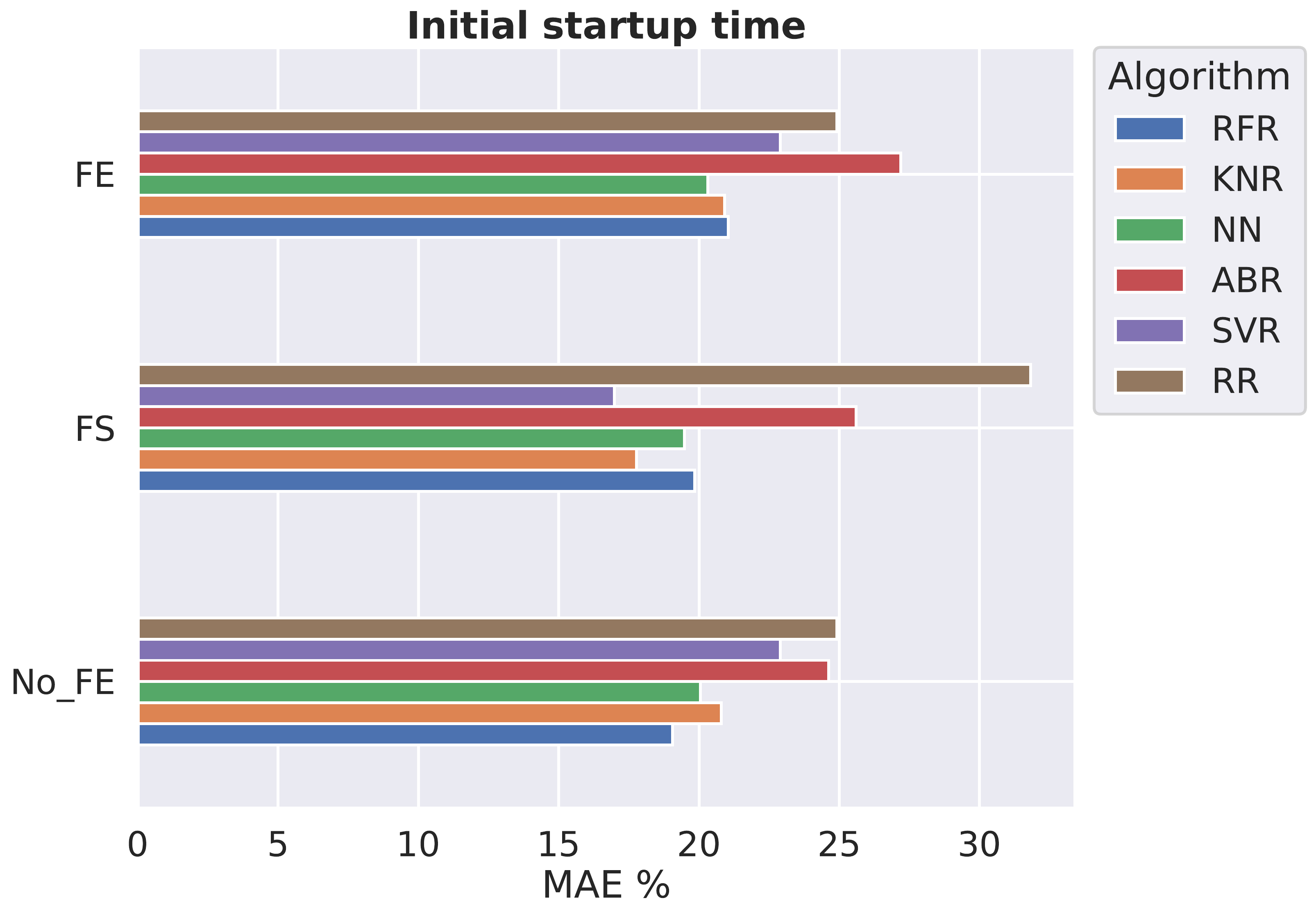}%
    \label{fig:initPlayingTimeMAE}
    }
\subfigure[Prediction time]{
    \includegraphics[width = 0.47\textwidth]{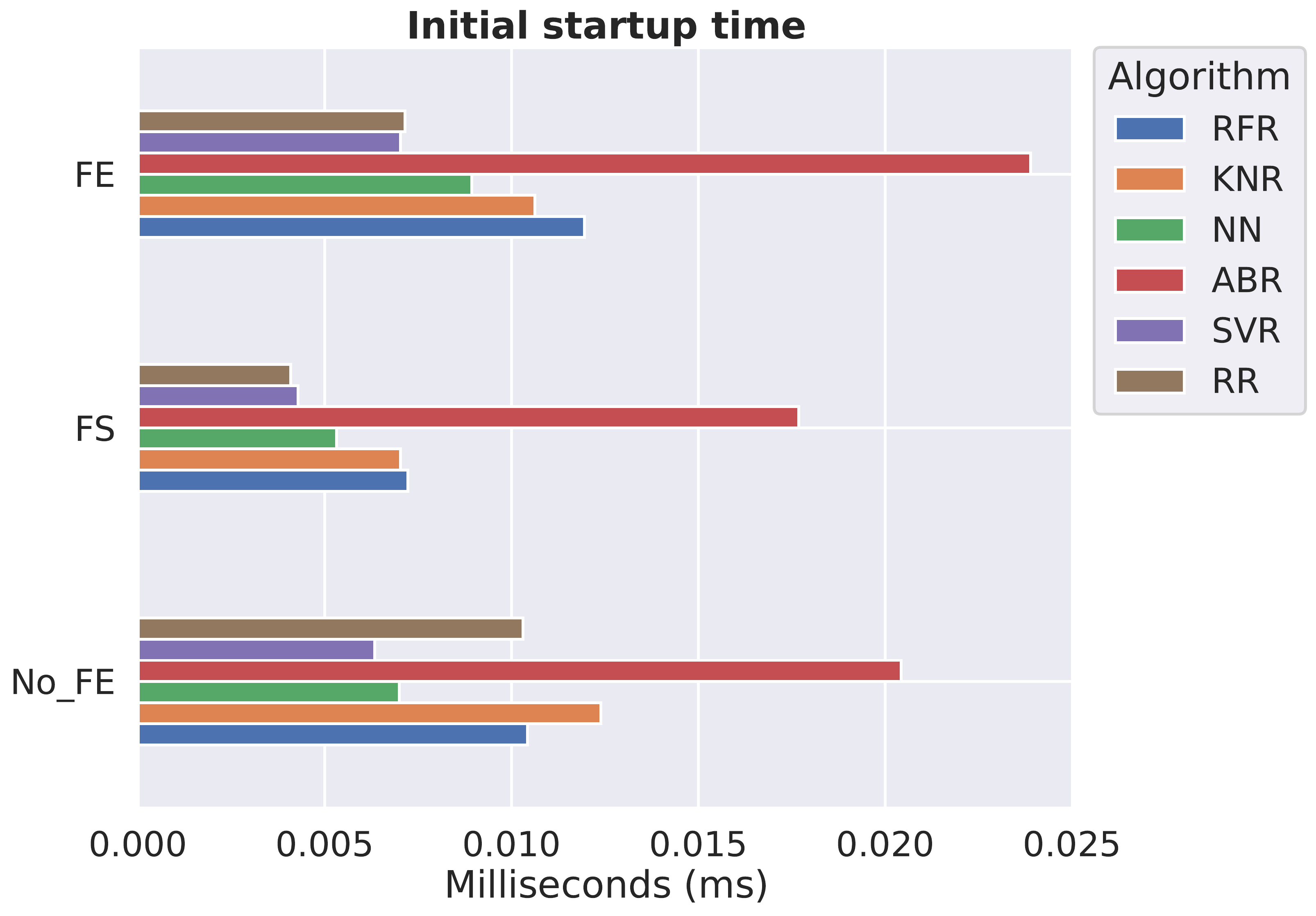}%
    \label{fig:initPlayingTimeptime}
    }
\caption{ML model performance for Initial startup time.}
\label{fig:initPlayingTime} 
\end{figure*}

In Figure \ref{fig:initPlayingTimeMAE} the MAE\% reached by each ML algorithm is depicted. As it can be seen, the algorithm that estimates the best for the \textit{init startup time} KQI is SVR using an FS approach. However, KNR shows good accuracy using the same strategy. The results reached by FE are clearly outperformed by FS with MAE values over 20\% of error. For instance, if the average value of the initial startup time in the dataset is 1s, a 20\% of error means that the estimation has an average error of $0.2$s. 

Resuming the analysis, it is remarkable the performance reached by the algorithms when no feature engineering technique is applied. This is especially notable with RF and NN (MLP) approaches.

Conversely, Figure \ref{fig:initPlayingTimeptime} describes the estimation time. As can be seen, the algorithms that perform the fastest estimations are RR and SVR for FE and FS. Nonetheless, the factor that contrasts the most is the estimation time reached by the algorithms using the FS approach. This shows that the FE approach fits perfectly for this metric, outperforming the other algorithms regarding both the quality and speed of the estimations. This is based on that this strategy generates a lower model complexity due to the use of fewer input variables, as well as no additional mathematical operations are required, like in the case of FE with the PCA transformation. Nonetheless, all the algorithms provide fast-enough estimations, in the order of milliseconds, of the averaged metrics per session, which present a coarse granularity.

\begin{figure*}[ht]
\centering   
\subfigure[Estimation error]{
    \includegraphics[width = 0.47\textwidth]{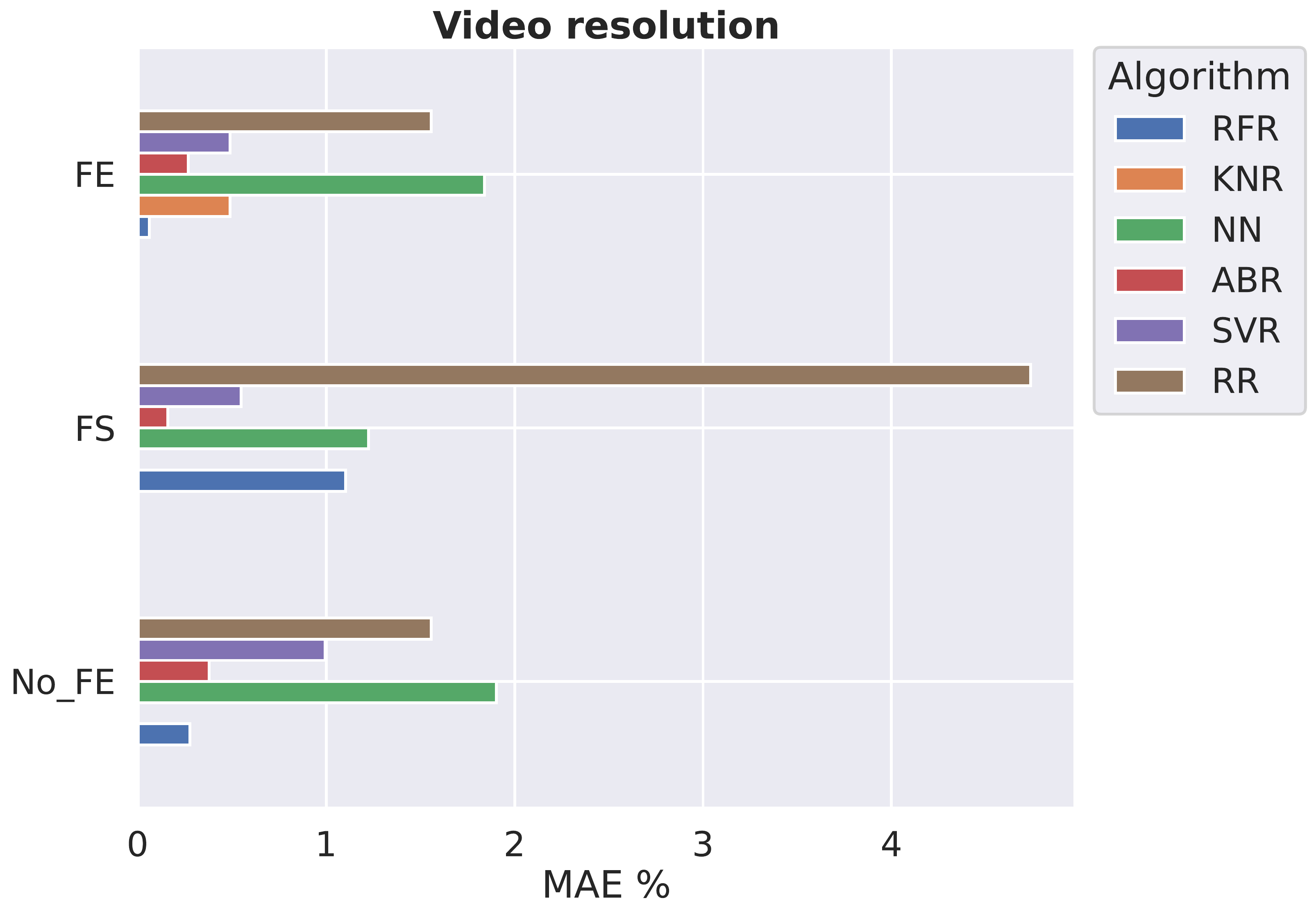}
    \label{fig:videoWidthMAE}
}
\subfigure[Prediction time]{
    \includegraphics[width = 0.47\textwidth]{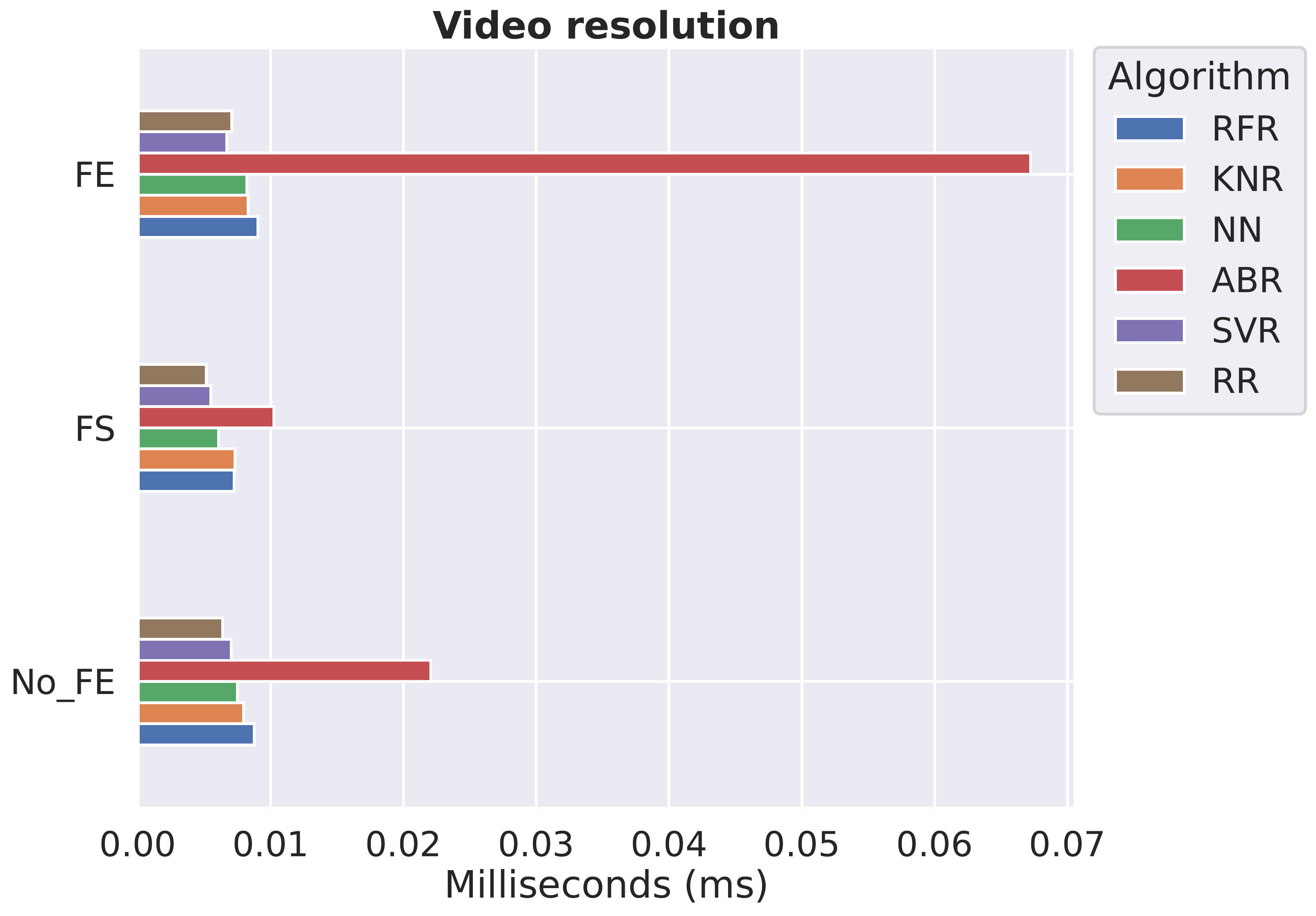}
    \label{fig:videoWidthptime}
}
\caption{ML model performance for Video resolution.}
\label{fig:videoWidth} 
\end{figure*}

One of the most important service metrics is video resolution, which provides salient insights into the quality of the experience. In order to represent the resolution, which is basically the number of rows by columns of pixels, this has been mapped into a one-digit scale (e.g. 0 for 0x0 resolution, 1 for 720x340, and so on and so forth). This decision is based on the finite number of resolutions possible, nonetheless, the results obtained are similar to just independently estimating rows or columns. As depicted in Figure \ref{fig:videoWidthMAE} all the algorithms perform remarkably well in terms of metric estimation. The best results are obtained with a KNR approach for FS and Non-FE reaching a MAE\% close to 0.00\%, which in essence indicates that this KQI is subject to precise estimation using network information. Despite this specific case, it is possible to observe that all the approaches reach less than 2.00\% of error with the exception of RR.

Likewise, Figure \ref{fig:videoWidthptime} shows the prediction time accounted for every algorithm. The results demonstrate a similar pattern to the initial startup time case, where performance homogeneity is the key characteristic for all the cases. Even though the ABR algorithm time seems to be higher in comparison with the other ones, the estimation is done in times less than 0.1 ms. This time is important taking into account its application within the process of scheduling in service-optimized networks. 

In this context, as a side-analysis, the estimation of the video resolution can develop an improvement in the estimation of other metrics correlated with its implicit information. For instance, a higher resolution implies more transport resources, which may lead to an increment in the probability to suffer from stalls in the playback or the rise in the initial playback time values.

Another KQI analyzed is the video frame rate displayed on the user side through the HMD. This parameter depends on the number of downloaded frames rather than on the hardware capacity of the device. As expected, the assessment has turned out in low MAE for most of the ML techniques and approaches used (i.e., No\_FE, FS, FE). This outcome conduct is related to the nature of the evaluated metric, which has low variability due to the \textit{Adaptive Bitrate} strategy adopted in the source in order to modify the resolution to maintain a stable frame rate. Given what has been said, the best results are obtained with KNR but with similar performance to RF. On the other hand, the estimation time exhibits an equivalent behavior to the previous cases, where the general trend is to have faster estimations with the FS strategy.

\begin{figure*}[ht]
\centering
\subfigure[Estimation error]{
    \includegraphics[width=0.47\textwidth]{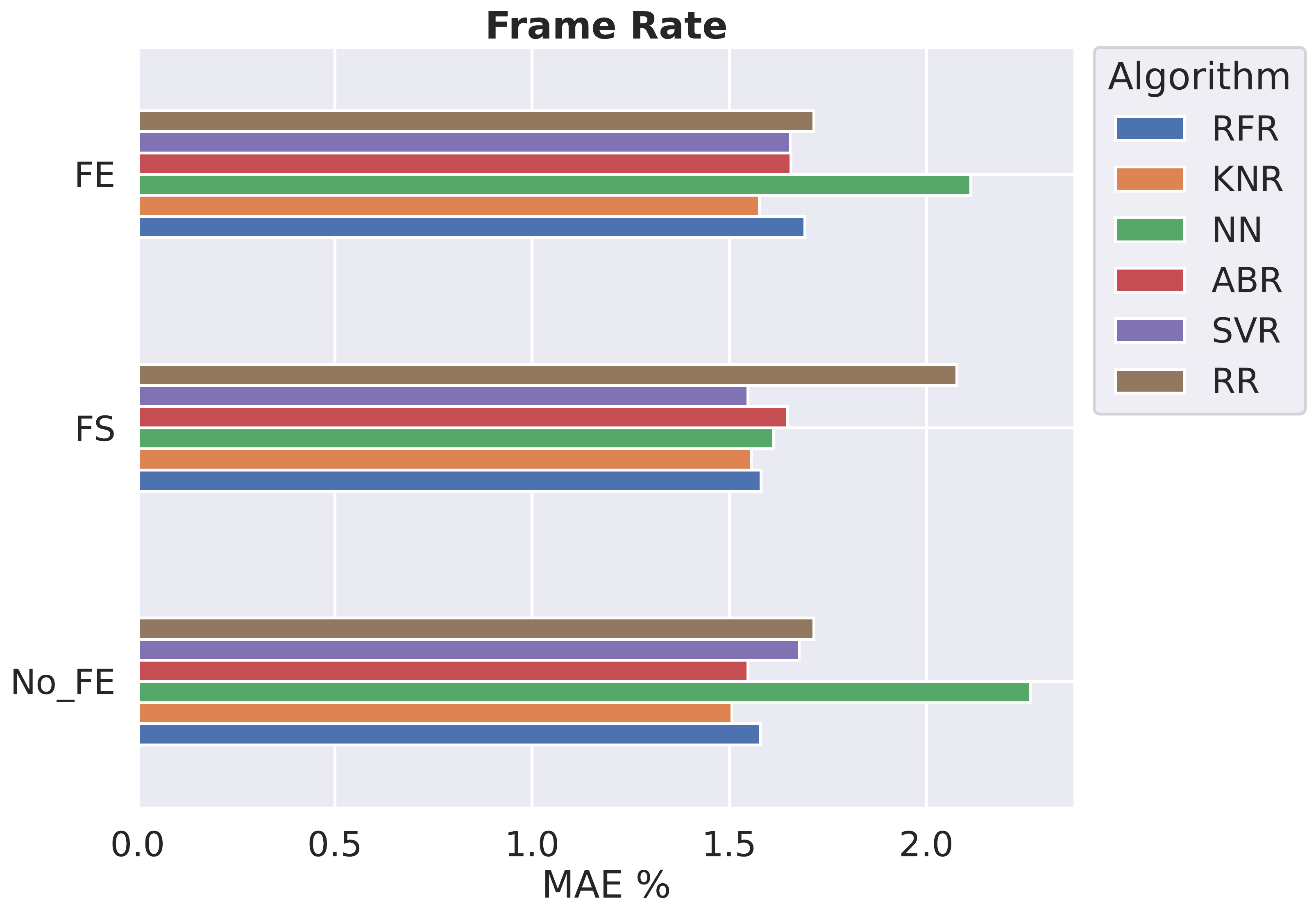}%
    \label{fig:displayRateMAE}
    }
\subfigure[Prediction time]{
    \includegraphics[width=0.47\textwidth]{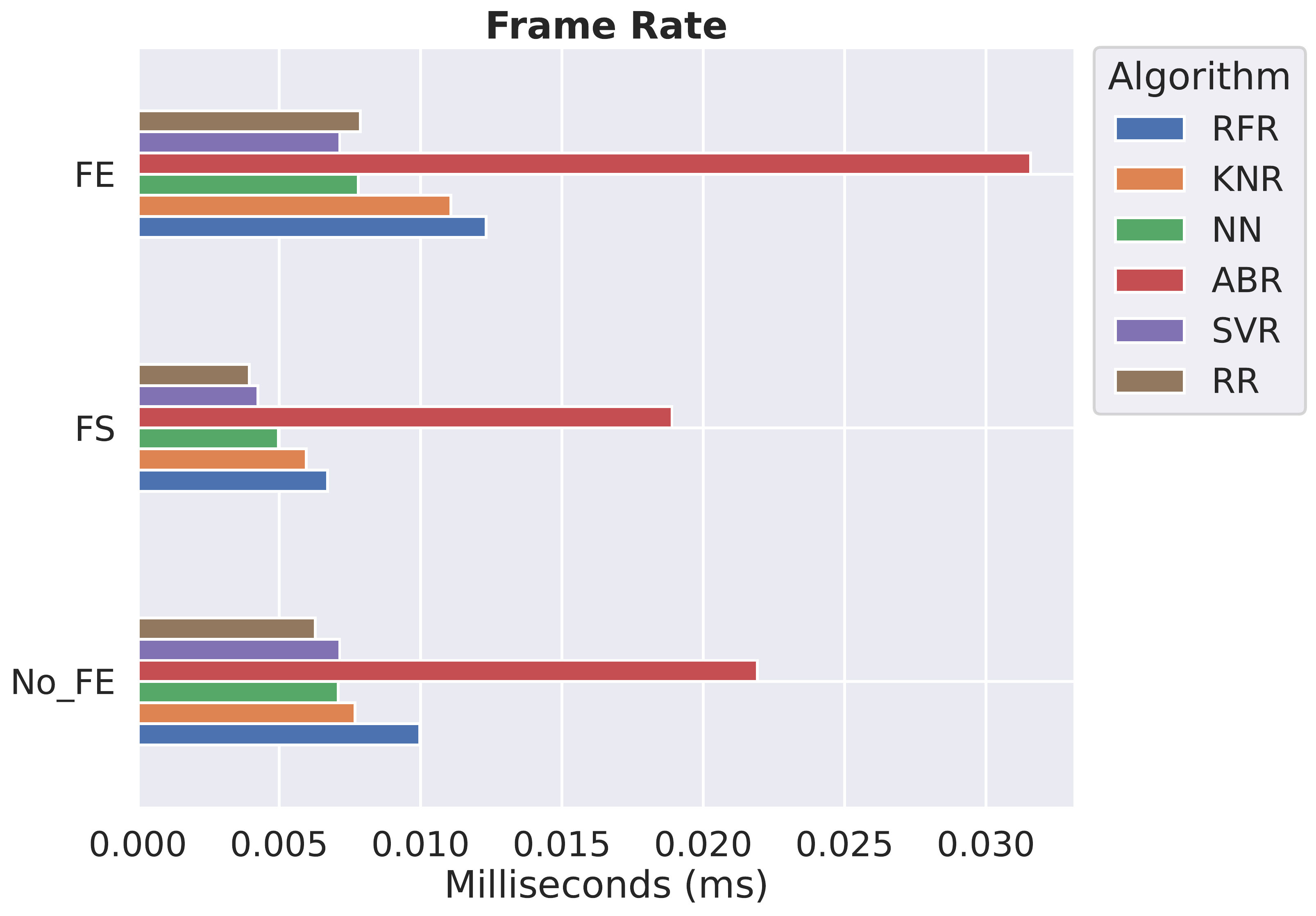}%
    \label{fig:displayRateptime}
    }
\caption{ML model performance for Frame Rate.}
\label{fig:videoDisplayRate} 
\end{figure*}

\begin{figure*}[ht]
\centering
\subfigure[Estimation error]{
    \includegraphics[width=0.47\textwidth]{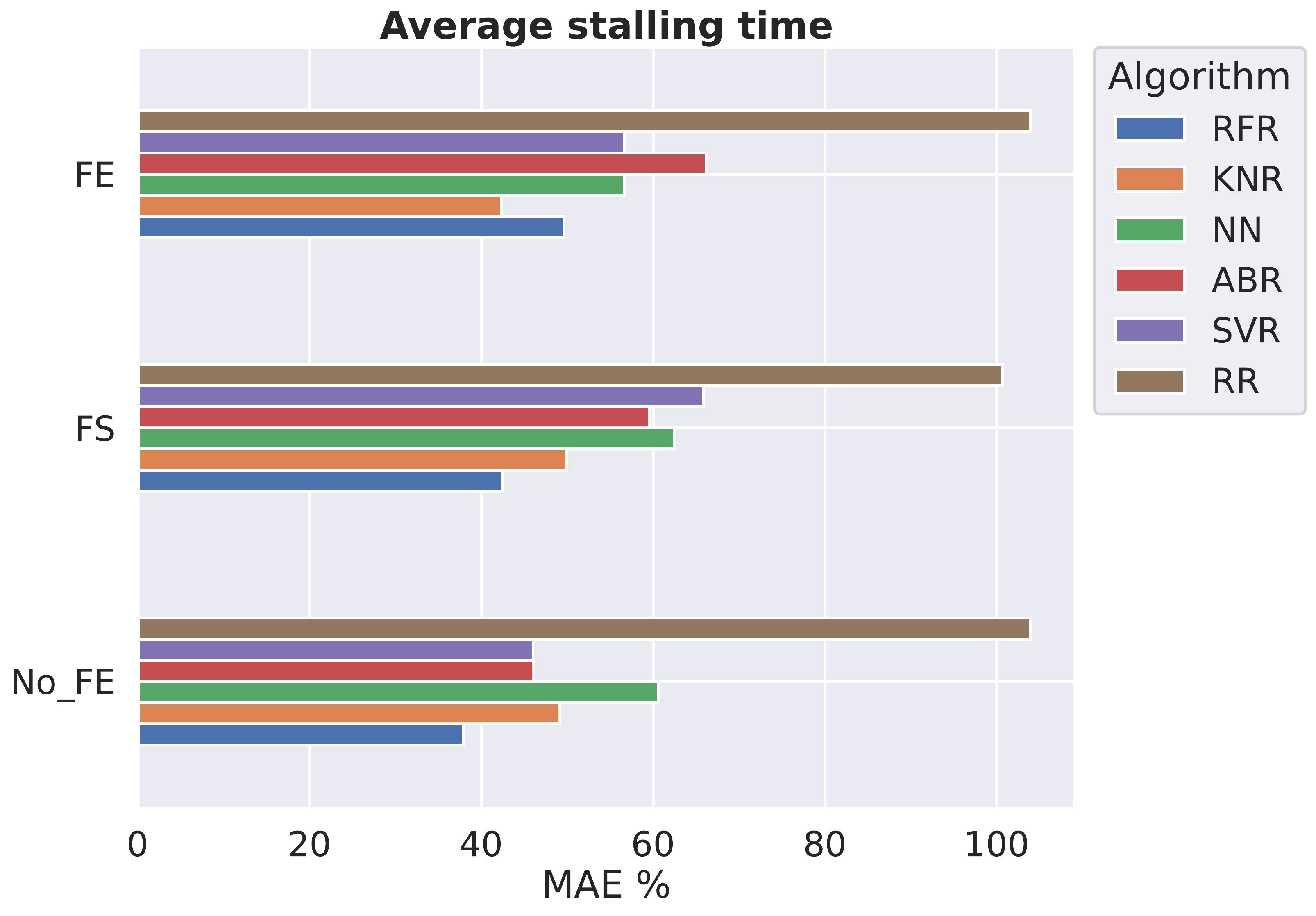}%
    \label{fig:avgStallTimeMAE}
    }
\subfigure[Prediction time]{
    \includegraphics[width=0.47\textwidth]{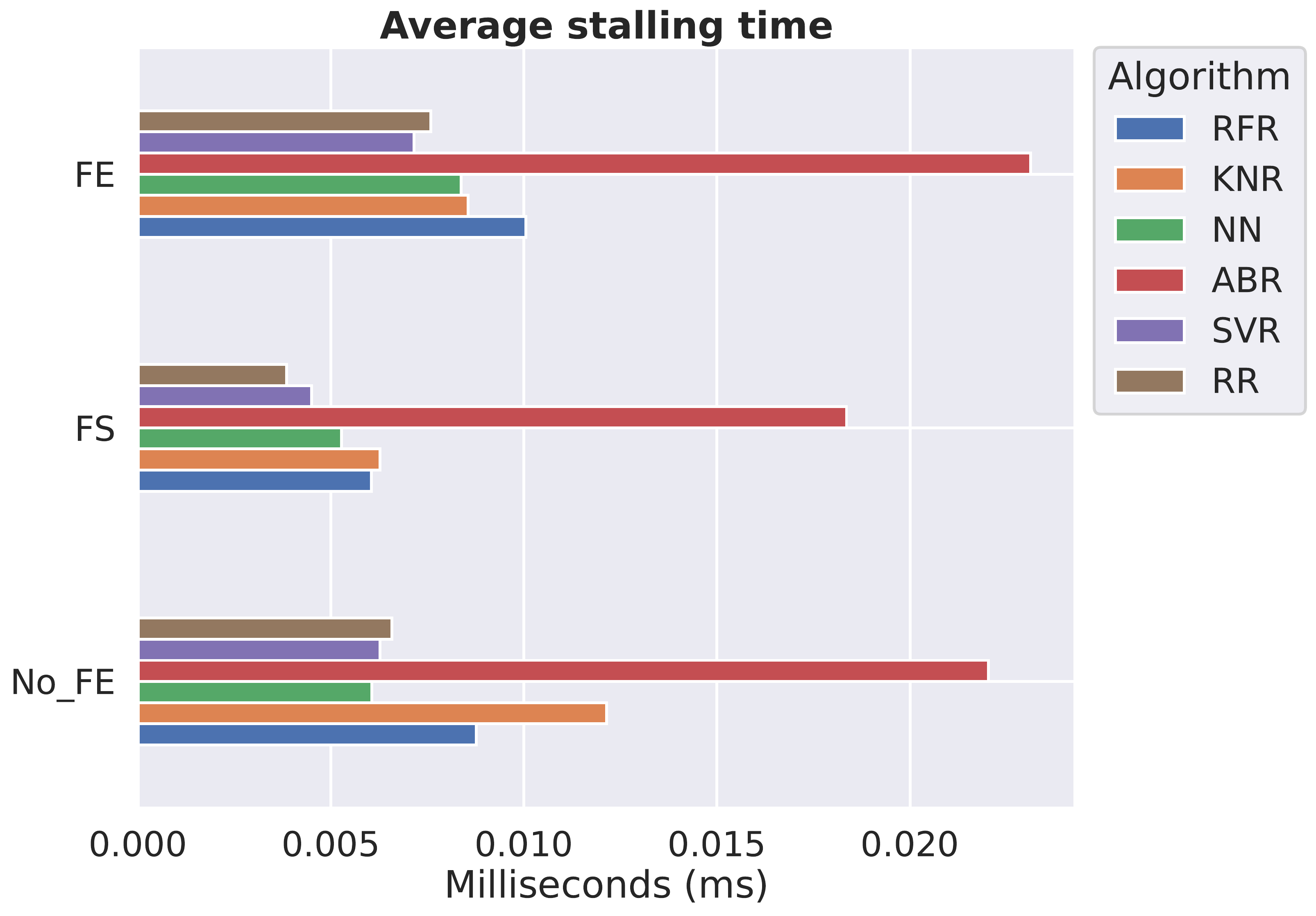}%
    \label{fig:avgStallTimeRMSE}
    }
\caption{ML model performance for Stalling time.}
\label{fig:avgStallTime} 
\end{figure*}

Contrary to the previous cases, the average stall time is a high-variance variable. Its values depend on parameters such as the video resolution, frame rate, the current network conditions as well as the radio ones. These effects can be seen in the results gathered for ML estimation in Figure \ref{fig:avgStallTimeMAE}. As shown, the MAE\% exhibits elevated inaccuracy of certain algorithms with special mention to RR, which has been outperformed by all the other algorithms in the previous analysis.  Despite this, the KNR approach, which has shown a proper performance for the last metrics presents a suitable performance close to 50\%. Nevertheless, RF presents the best performance for this case with a MAE\% of about 40\%. In this sense, the performance of these models seems not to be suitable, as in the previous cases. This is because the stalling time is a quite difficult variable to predict because of its dependency on multiple factors. Nonetheless, it is valuable to get an insight into this metric which is helpful for future network optimization.

Following the thread of this metric, the estimation time provides a similar point of view, where the ABR algorithm requires more time to predict the metric for all the approaches. Unlike the estimation time case, results also show that the use of FE or FS techniques in this KQI (i.e., stalling time) does not add any advantage with regard to direct approaches.

\begin{figure*}[ht]
\centering
\subfigure[Estimation error]{
    \includegraphics[width=0.47\textwidth]{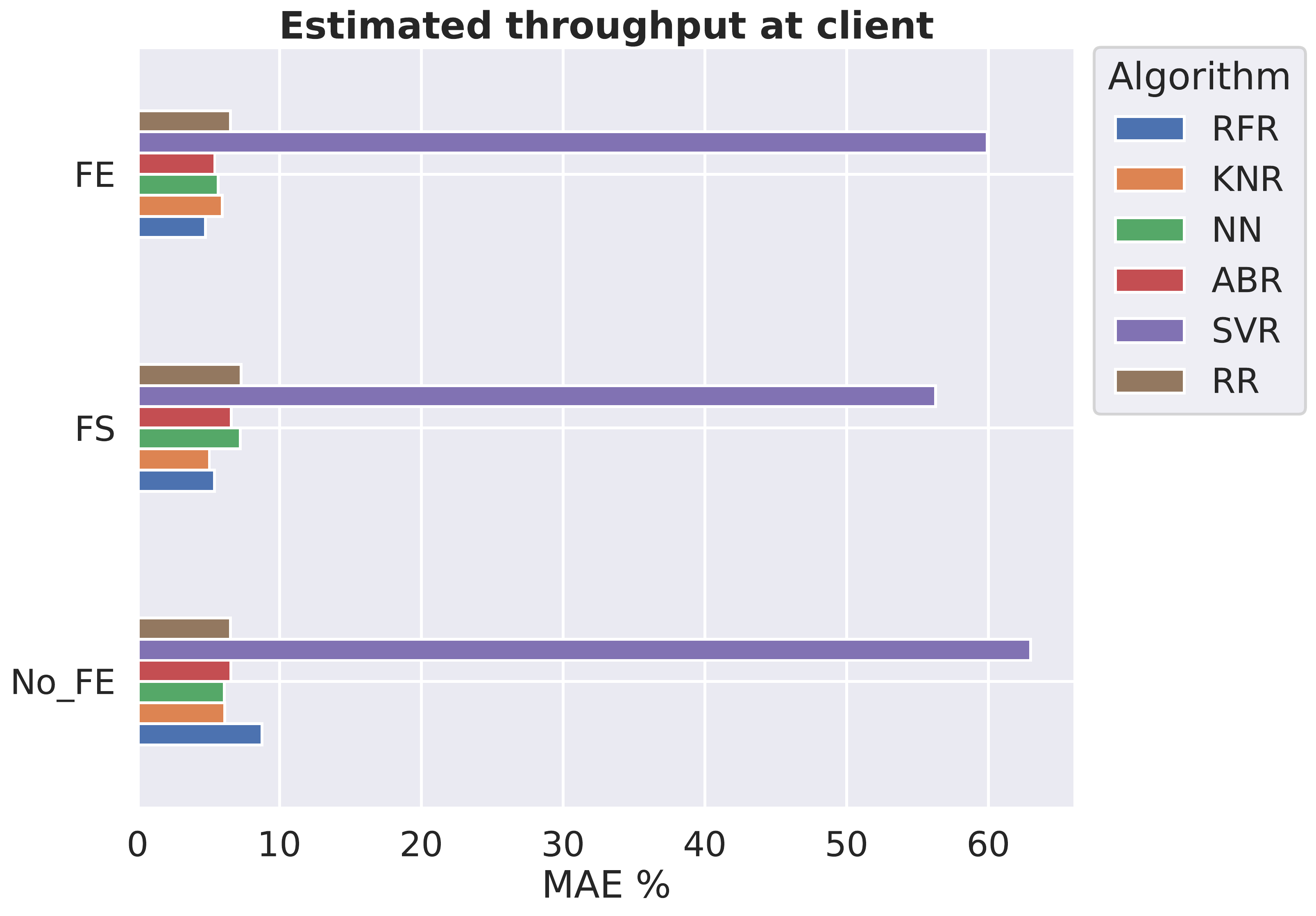}%
    \label{fig:throughputMAE}
    }
\subfigure[Prediction time]{
    \includegraphics[width=0.47\textwidth]{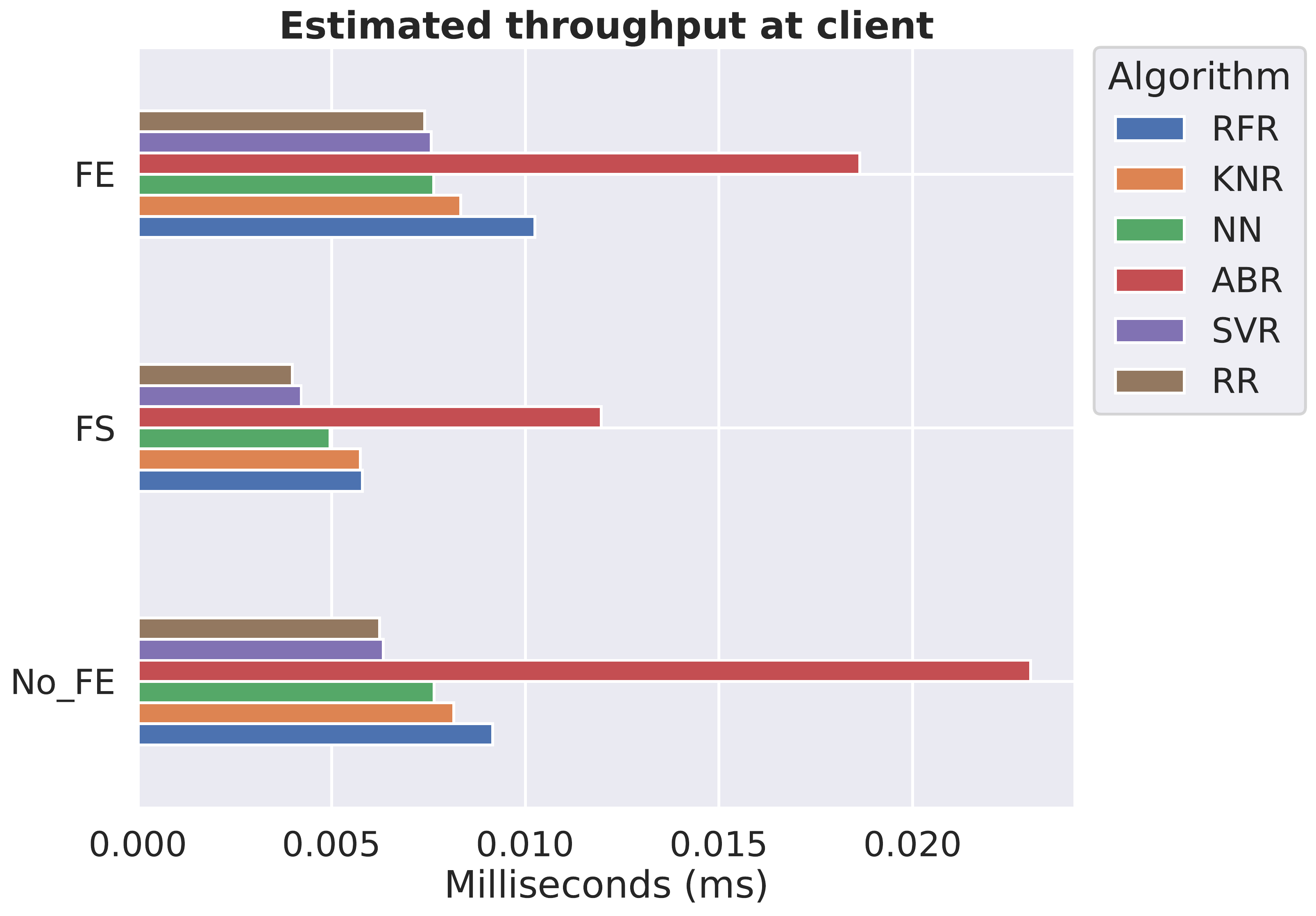}%
    \label{fig:throughputptime}
    }
\caption{ML model performance for Throughput.}
\label{fig:throughput} 
\end{figure*}

On its behalf, the throughput estimated at the client side is an important metric that reflects a quality-of-the-service viewport based on the approximate quantity of information that arrives at the device. In this context, this parameter may affect directly the other parameters involved in this work. A constrained throughput can carry to a low-resolution video service, or alternatively, lead to the increase of stalling events or the startup time of this service. For this purpose, this parameter has been assessed as well as the others. The results in Figure \ref{fig:throughputMAE} report that the KNR algorithm performs the best among the others with a MAE\% of about 5\%. However, it is remarkable that all the other algorithms, except for SVR, perform very acceptable (less than 9\%). When it comes to the estimation time, the results show the same pattern exhibited for the other estimated metrics.

\begin{figure*}[ht]
\centering
\subfigure[Estimation error]{
    \includegraphics[width=0.47\textwidth]{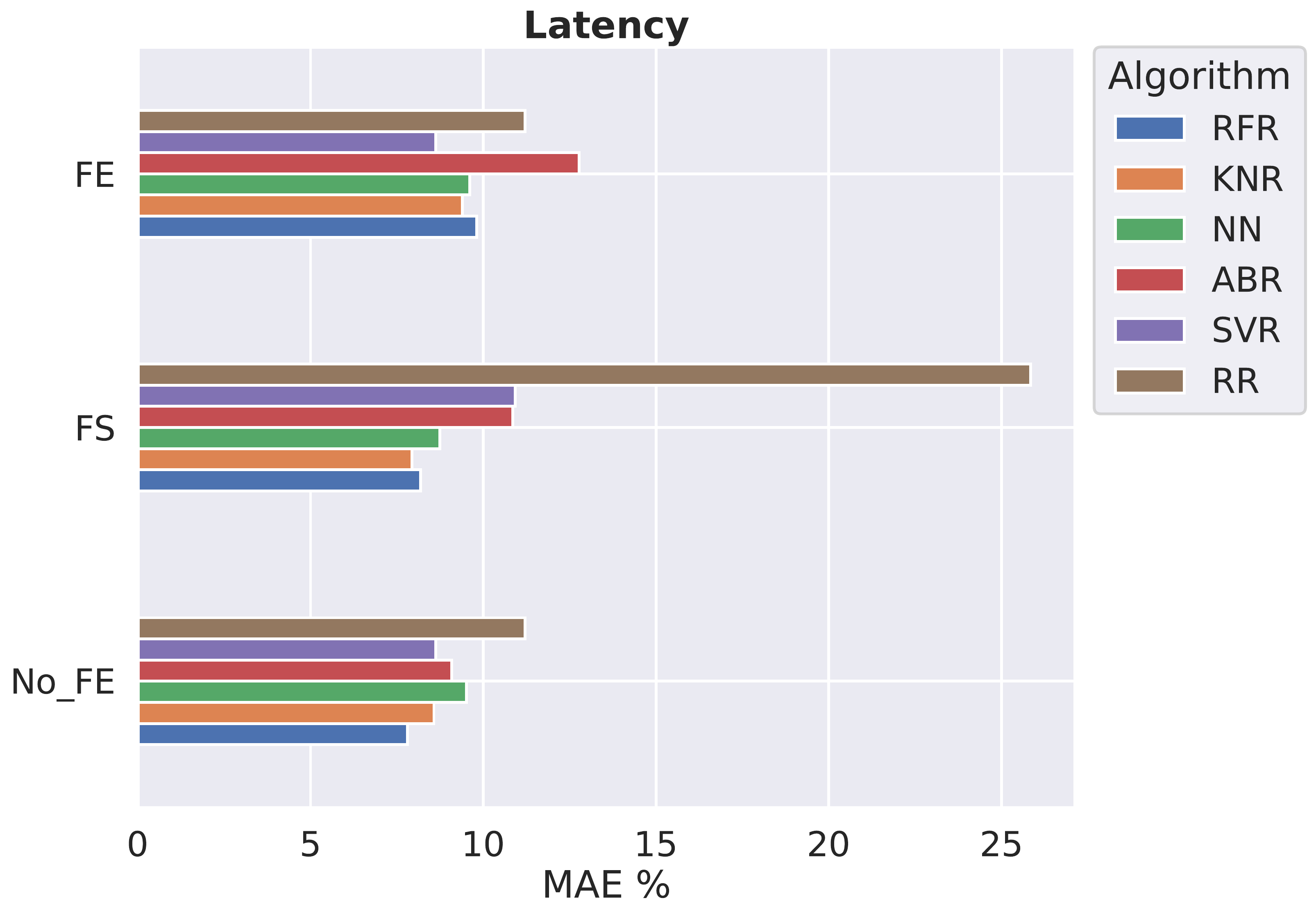}%
    \label{fig:rttMAE}
    }
\subfigure[Prediction time]{
    \includegraphics[width=0.47\textwidth]{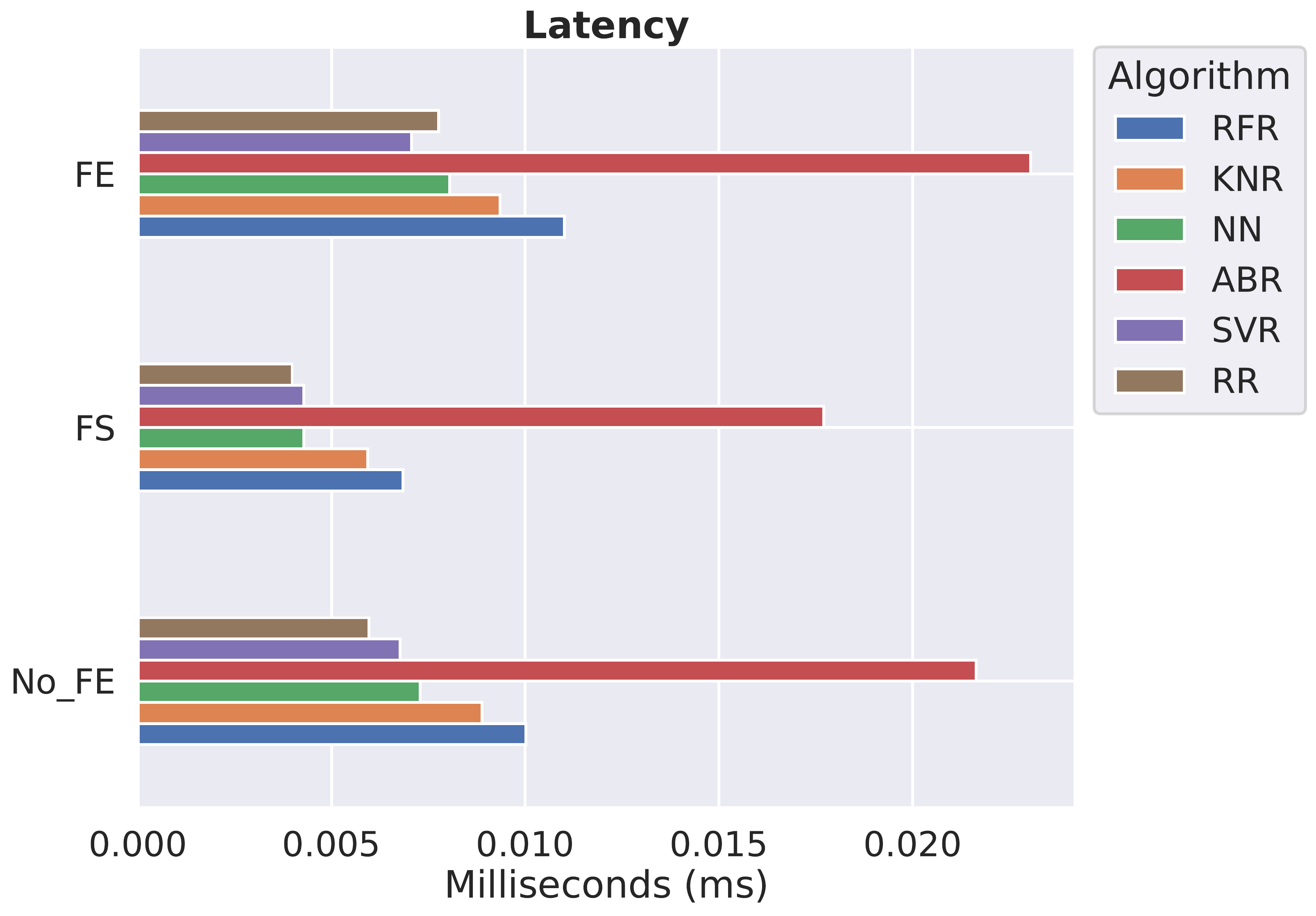}%
    \label{fig:rttptime}
    }
\caption{ML model performance for Latency.}
\label{fig:rtt} 
\end{figure*}

Finally, the last KQI analyzed is the latency between the HMD and the video server, considering both sides of the communication. As can be seen in Figure \ref{fig:rttMAE}, the obtained results show good performance for most of the algorithm, except RR. Likewise the other estimation cases, FS presents the best approximations compared with the ground-truth values in the dataset. In this setting, KNR using FS outperforms the other options with a MAE\% of about 7.5\%. This is a remarkable value due to the difficulty to estimate a real E2E latency from a service perspective. In addition, the knowledge of this metric can provide key insights into other KQIs like stalling events, initial startup time, etc. In this context, the latency can give an adequate perception of the level of the network stress which produces effects on the service quality experience.

\begin{figure}[ht]
    \centering
    \includegraphics[scale = 0.3]{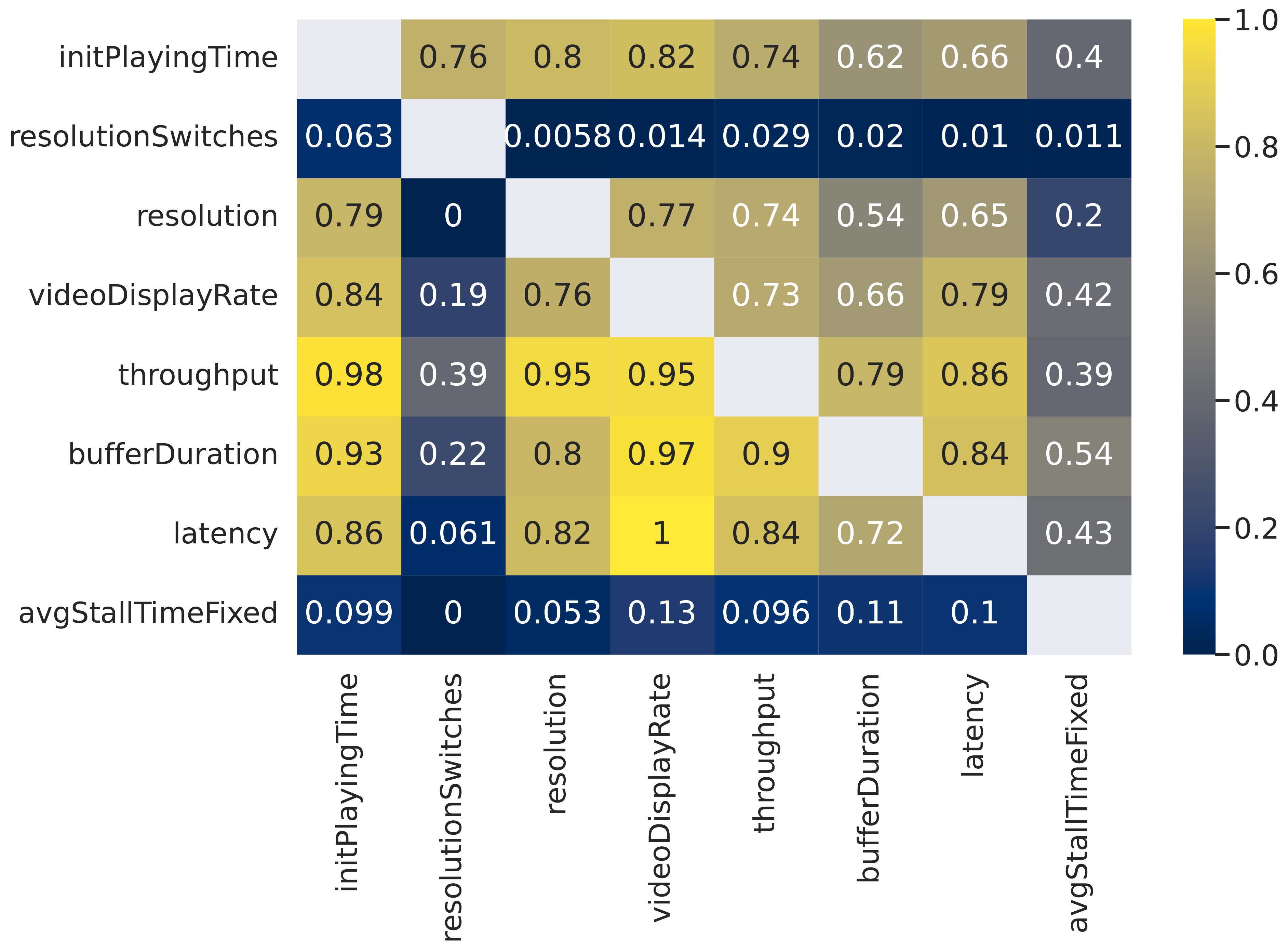}
    \caption{Mutual information between KQIs}
    \label{fig:MI}
\end{figure}

To support the fact that certain KQIs are more difficult to estimate with respect to others, a matrix showing the shared information between KQIs is displayed in Figure \ref{fig:MI}. The strategy adopted to evaluate the relations among these metrics is the calculation of the Mutual Information (MI) between KQIs. For instance, the resultant matrix shows that the throughput can highly impact the initial playing time, the video resolution, the frame rate, and the latency, which is logical as previously explained in the throughput estimation analysis. Conversely, the MI between the stalling time with the other KQIs is almost negligible. In this context, this means that is difficult to estimate this variable from other high-layer indicators as well as from network-based metrics, incurring higher MAE\% metrics in comparison with other ML models.

\begin{figure*}[ht]
    \centering
    \includegraphics[trim = 0 0 0 0, clip, width = 0.65\textwidth]{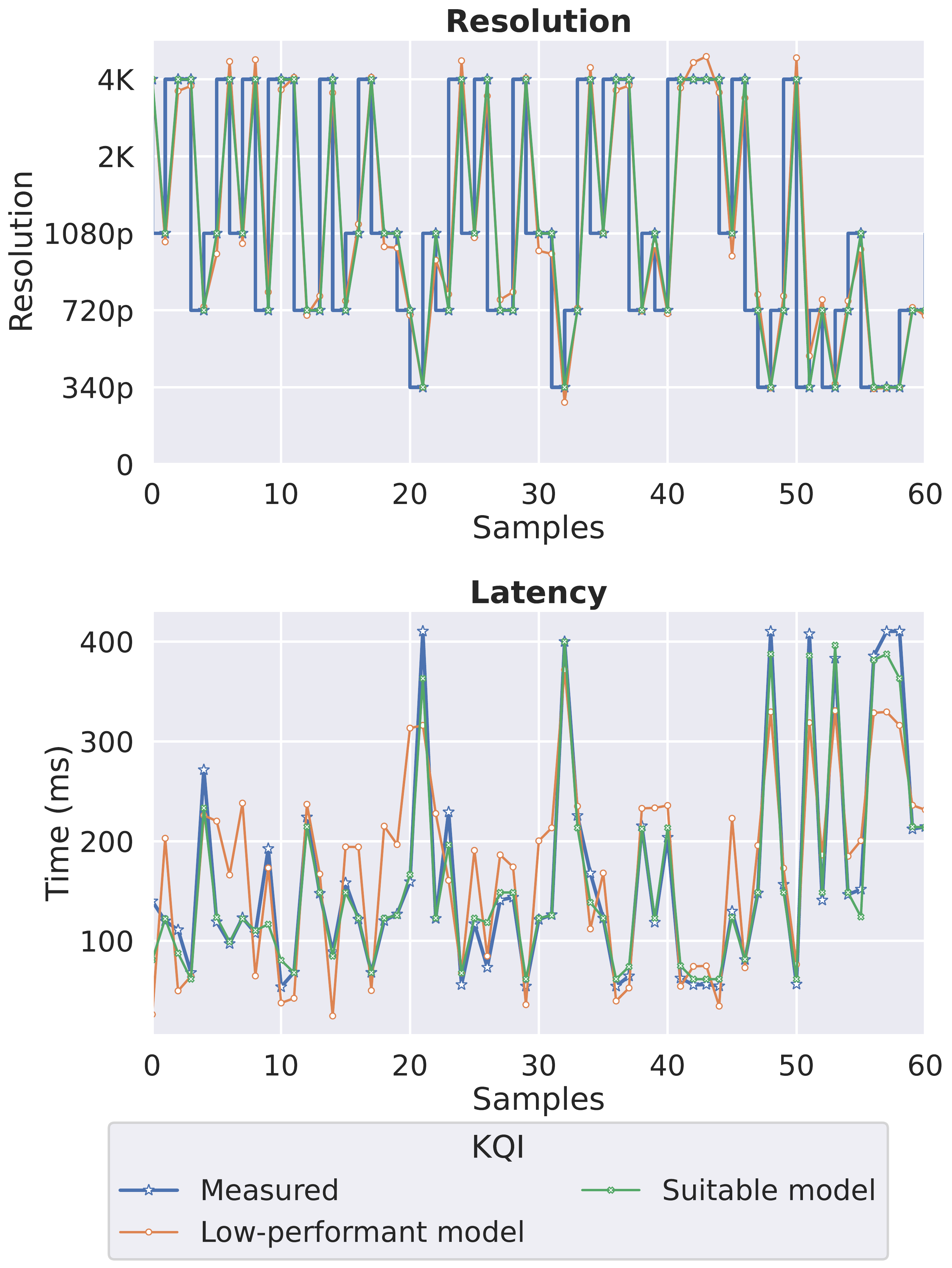}
    \caption{Estimation example.}
    \label{fig:estimation_example}
\end{figure*}

A graphical representation of the capacity of this ML framework is depicted in Figure \ref{fig:estimation_example}. This illustration shows the estimation of resolution and latency KQIs in contrast with the ground truth values. The subfigure at the top shows the estimation performance for the video resolution KQI, while the latency is at the bottom. In both cases, it is presented models with different levels of performance. As can be seen, a suitable model is able to predict KQIs with high precision from the own information from the network, which denotes the importance of the application of ML-based approaches in the management of mobile networks.

To summarize, this section has presented some insights into the use of ML to estimate high-layer metrics which are significant in the process of network optimization based on E2E QoE. The collected outcomes expose proper performance for some of the algorithms assessed, albeit, it is remarkable the performance of KNR in most of the cases. Furthermore, three different approaches were evaluated in order to improve the process of modeling and estimation such as feature extraction, feature selection, and without any feature engineering strategy. The results exhibit a common pattern: FS endows the models with less error, and in most cases, the least estimation times, because of its reduced model complexity. This is a point that should be considered taking into account that ML can provide a remarkable improvement in the way resources are managed by networks in order to provide high-quality services. 

\section{CONCLUSION}
\label{sec:Conclusion}

This work has presented an ML approach for KQI estimation in the field of XR services. For this purpose, a 360-Video service has been selected as a use case. This ML-based framework aims to automate the process of fetching objective quality metrics for QoE assessment. In this context, this strategy exploits the information self-contained in network metrics and configuration parameters that are reachable for network operators. The key advantage of this approach is the ability to speed up the process of service management from an E2E perspective, as well as introduce new features that may be used to improve the network performance in terms of the service experience. 

In this sense, this work described the process to generate the input dataset using an LTE/5G 360-Video testbed, which captures the network impact on the service using different experiments with diverse configurations of transmission power, noise, and channel bandwidth. Then, the exploitation of this information is relevant for the model training phase. In addition, feature engineering techniques such as feature selection and feature extraction have been included to establish which mechanism performs better for this kind of data. Furthermore, the framework implemented and assessed different ML algorithms through a search-grid scheme to determine the best hyperparameter tuning for each model using a 5-fold strategy. The performance assessment is done through the Scaled Mean Average Error (MAE\%) and the estimation time.

Results have shown that suitable performances are obtained for KQI estimation in the context of 360-Video. Regarding this, KNR outperforms the other algorithms by showing reduced MAE\% in most cases. In this setting, RF arises as a possible alternative to the first one. Nonetheless, the way each algorithm works is affected by the nature of the metrics. The average stall time and initial playback time, on a smaller scale, are variables with high variance impacting negatively where the samples are outliers. Conversely, metrics like resolution and frame rate, and even throughput allowed the algorithms to satisfactory perform along their estimations. This fact was supported by the use of an MI matrix, which depicts the relations between KQIs, exhibiting the same behavior.

As a future research line, the impact of nested estimation of the metrics may be performed in order to enhance the accuracy of the algorithms. The exploitation of previously estimated KQIs may be useful to strengthen the statistical information in the training set, according to the MI matrix, this way reducing the MAE of the estimations. Moreover, it is planned to work on the implementation of network configuration mechanisms to improve its performance by exploiting some 5G/B5G enabler technologies such as network slicing, virtualization, MEC, etc. In addition, it is possible to explore this strategy oriented to its application on novel network architectures like Open RAN (e.g. x-App and r-App design).

 \bibliographystyle{elsarticle-num-names} 
 \bibliography{Bib}

\end{document}